\documentclass[aps,prl,reprint,superscriptaddress]{revtex4-2}
\usepackage{chemformula} 
\usepackage[T1]{fontenc} 
\usepackage{graphicx} 
\usepackage{dcolumn} 
\usepackage{bm} 
\usepackage[colorlinks, linkcolor=blue, citecolor=blue, urlcolor=blue]{hyperref}

\begin{document}
\title{Partial-Wetting Phenomena in Active Matter}

\author{Jing Zhang}
\affiliation{School of Physical Science and Engineering, Beijing Jiaotong University, Beijing 100044, China}
\affiliation{Beijing Key Laboratory of Novel Materials Genetic Engineering and Application for Rail Transit, Beijing Jiaotong University, Beijing 100044, China}

\author{Zhixin Liu}
\affiliation{School of Physical Science and Engineering, Beijing Jiaotong University, Beijing 100044, China}
\affiliation{Beijing Key Laboratory of Novel Materials Genetic Engineering and Application for Rail Transit, Beijing Jiaotong University, Beijing 100044, China}

\author{Shengda Zhao}
\affiliation{School of Physical Science and Engineering, Beijing Jiaotong University, Beijing 100044, China}
\affiliation{Beijing Key Laboratory of Novel Materials Genetic Engineering and Application for Rail Transit, Beijing Jiaotong University, Beijing 100044, China}

\author{Yangjun Yan}
\affiliation{School of Science, Key Laboratory of High Performance Scientific Computation, Xihua University, Chengdu 610039, China}

\author{Rongxin Yue}
\affiliation{School of Physical Science and Engineering, Beijing Jiaotong University, Beijing 100044, China}
\affiliation{Beijing Key Laboratory of Novel Materials Genetic Engineering and Application for Rail Transit, Beijing Jiaotong University, Beijing 100044, China}

\author{Jiaxin Yu}
\affiliation{School of Physical Science and Engineering, Beijing Jiaotong University, Beijing 100044, China}
\affiliation{Beijing Key Laboratory of Novel Materials Genetic Engineering and Application for Rail Transit, Beijing Jiaotong University, Beijing 100044, China}

\author{Xinjie Li}
\affiliation{School of Physical Science and Engineering, Beijing Jiaotong University, Beijing 100044, China}
\affiliation{Beijing Key Laboratory of Novel Materials Genetic Engineering and Application for Rail Transit, Beijing Jiaotong University, Beijing 100044, China}

\author{Xinghua Zhang}
\email{zhangxh@bjtu.edu.cn}
\affiliation{School of Physical Science and Engineering, Beijing Jiaotong University, Beijing 100044, China}
\affiliation{Beijing Key Laboratory of Novel Materials Genetic Engineering and Application for Rail Transit, Beijing Jiaotong University, Beijing 100044, China}

\begin{abstract}
Abundant interfacial phenomena in nature, such as water droplets on lotus leaves and water transport in plant vessels—originate from partial-wetting phenomena, which can be well described by Young's equation. It remains an intriguing question whether similar behaviors exist in active matter. In this letter, we present a clear demonstration of the partial-wetting phenomenon in a ternary laning system, which is a typical active system. A phase diagram is constructed in which the relative drift velocities of different components govern the transitions among drying, partial wetting, and complete wetting states. The mechanical balance on the contact lines of the partial-wetting phase described by Young's equation is verified. A theoretical picture is proposed to explain the analogy of partial wetting in the laning system to that in the equilibrium system. 

\end{abstract}
\maketitle


Wetting refers to the complete spreading of a liquid on a surface (contact angle $\theta\approx 0^\circ$), driven by the dominance of solid-liquid interfacial energy. Partial wetting occurs when the liquid forms a droplet with a finite contact angle ($\theta\in (0^\circ,180^\circ)$), governed by a balance of solid-liquid-gas interfacial tensions, where the contact angle and interfacial tensions obey Young's equation. Drying describes the complete removal of liquid from the surface\cite{SoftMatterPhysics2013,safranStatisticalThermodynamicsSurfaces2018,degennesWettingStaticsDynamics1985}.
Partial wetting is essential in natural biological systems, such as the self-cleaning properties of lotus leaves and the water-harvesting characteristics of cacti. Biomimetic superhydrophobic and superhydrophilic surfaces enable applications including self-cleaning glass, oil-water separation, and anti-biofouling interfaces\cite{yaoApplicationsBioInspiredSpecial2011,liuMultifunctionalIntegrationBiological2011}.

Active matter, a distinct class of nonequilibrium soft matter systems capable of consuming energy to generate self-propelled motion\cite{stenhammarActivityInducedPhaseSeparation2015,huberEmergenceCoexistingOrdered2018}, has emerged as a significant research frontier in recent years \cite{wysockiCapillaryActionScalar2020,siebertCriticalBehaviorActive2018}. Investigating interface tension-driven partial wetting in active matter systems poses an intriguing question. Recent studies in motility-induced phase separation (MIPS) systems have observed wetting-drying transitions around a permeable wall \cite{chaconIntrinsicStructurePerspective2022,klamserThermodynamicPhasesTwodimensional2018,catesMotilityInducedPhaseSeparation2015}. When the activity parameter exceeds a critical threshold, the dense phase liquid forms a continuous film on the walls, exhibiting complete wetting. Remarkably, increasing the particle permeability—achieved by tuning the wall porosity or reducing the energy barrier for particle penetration—induces a dynamic transition where the homogeneous wetting film destabilizes into isolated droplets, revealing a wetting-to-drying transition \cite{turciWettingTransitionActive2021a,dasMorphologicalTransitionsActive2020a,rojas-vegaWettingDynamicsMixtures2023}. Although the wetting-to-drying transition has been observed, there is still no direct evidence for partial wetting phenomena or quantitative validation of Young's equation in active matter. Partial-wetting behavior is governed by a balance of interfacial tensions at the contact line. However, for active matter, especially MIPS, there is as yet no agreed definition of surface tension. At low activity, active particles undergo conventional bulk phase separation. At higher activity, Ostwald ripening, the classical diffusive pathway to macroscopic phase separation, can go into reverse. \cite{tjhungClusterPhasesBubbly2018a,caballeroInterfaceDynamicsWet2025a,hermannNonnegativeInterfacialTension2019,faustiCapillaryInterfacialTension2021}. In addition, to observe partial wetting, the interfacial tensions among the three phases need to be well controlled. In studies of the wetting-drying transition of MIPS, although the solid-liquid and solid-gas interfacial tensions are adjustable, the liquid-gas interfacial tension remains inherently challenging to modulate. Consequently, it is hard to study partial wetting of active matter in single-component systems like MIPS.

This paper aims to investigate the partial wetting directly in nonequilibrium active systems and validate the mechanical balance at the contact line that governs this phenomenon. A multiphase system formed through phase separation in an A-B-C ternary laning system is employed, facilitating enhanced observation of partial wetting behavior compared to single-component systems like MIPS. The laning phenomenon, as one of the simplest models characterizing nonequilibrium phase separation in active matter, is typically studied in binary systems \cite{sutterlinDynamicsLaneFormation2009}, where two components undergo counter-directional drift motions \cite{reichhardtLaningClusteringTransitions2018}. When the driving force exceeds a critical threshold, the system self-organizes into lanes aligned with the external field direction \cite{bacikLaneNucleationComplex2023,poncetUniversalLongRanged2017}. This self-organization phenomenon manifests ubiquitously across diverse systems, including pedestrian traffic dynamics \cite{joseSelfOrganizedLaneFormation2020}, molecular motor transport along microtubules, colloidal suspensions \cite{duExperimentalInvestigationLane2012,killerPhaseSeparationBinary2016}, electrolytes, plasmas, ionic liquids, and vibrated granular systems \cite{mullinCoarseningSelfOrganizedClusters2000}. This study constructs a ternary laning system using the differences in coupling coefficients $q_{\alpha}$ between the component $\alpha=A, B, C$ and the external field, to achieve three coexisting lane-ordered phases. At the interface formed by A-B phase separation, the compatibilities of component C with component A and B are modulated by $q_C$ and then the interfacial tensions are controlled, thereby enabling observation of partial wetting phenomena. A quantitative relationship between macroscopic interfacial tensions and the microscopic coupling parameter $q_\alpha$ is constructed. This enables quantitative validation of Young's equation in active matter systems.

Consider a laning system that consists of three components of spherical particles, A, B, and C, with the total number of particles given by $N = \sum_\alpha N_\alpha$. Here, $N_{\alpha}$ represents the number of particles of the component $\alpha=A,B,C$. Mutual interactions are governed by the potential of Weeks-Chandler-Anderson (WCA). The energy and length scales are $k_BT$ and $\sigma$, respectively \cite{dzubiellaLaneFormationColloidal2002,rednerStructureDynamicsPhaseSeparating2013b,xuMorphologiesDynamicsInterfaces2021,xuMorphologiesDynamicsInterfaces2021}. The particles are driven by an external field $\mathbf{E}$. The force acting on the $i$-th particle is given by  
$
\mathbf{f} = \mathbf{E} q_\alpha.
$
The coupling coefficient $q_\alpha \in [-1,1]$ can also be interpreted as the "charge" of the particle, its magnitude controls the strength of the force, while its sign determines the direction of the force. The laning structure will be formed along $\mathbf E$, which is from a lateral phase separation perpendicular to $\mathbf E$. Generally, laning behaviors were studied in 2D systems. However, partial wetting involves an inhomogeneous distribution of component C along the A-B interface, which requires at least a 1D interface in the lateral direction. Therefore, the simulations are performed in a 3D space.  
The direction of $\mathbf E$ is defined as the $z$-axis. The normal direction to the A–B interface is along the $y$-axis, while the interface itself extends along the $x$-axis. Any arbitrary vector perpendicular to $\mathbf E$ is represented by $\mathbf{R}$. The average number density of particles is fixed at $\bar\rho = 0.5$ (for more details on the model, parameters, and simulation methods, see Sec.~I of the Supplemental Material (SM) \cite{supplemental1}.).

Before presenting the partial wetting, we first demonstrate the phase separation and the interfacial properties of a binary laning system in 3D space, and attempt to elucidate the relationship between the interfacial tension $\gamma$ and the microscopic parameters $E$ and $q_\alpha$. The system is initialized with a randomly mixed configuration and evolves to a steady state. For a sufficiently large field, phase separation takes place, which can be characterized by the order parameter\cite{dzubiellaLaneFormationColloidal2002}:  
$
\psi = \frac{1}{n}\left\langle\sum_{i=1}^n \left[\frac{n_a(i) - n_b(i)}{n_a(i) + n_b(i) }\right]^2\right\rangle.
$ 
Here, $ n_a(i) $ and $ n_b(i) $ denote the numbers of neighboring particles within a lateral distance of $ 0.75\sigma $ with the same and different species with the $i$ th particle, respectively. $ \psi \to 0 $ in a homogeneous state, while $ \psi \to 1 $ indicates phase separation. Figure~\ref{fig:fig1}(a) shows the dependence of $ \psi $ on $ \Delta q E $ for different values of $ \Delta q = q_A - q_B $
under $ \phi_0 = 0.5 $. All data collapse onto a single universal curve, which indicates that $ \Delta q E $ acts as the driving force for phase separation.
The phase separation point can be identified by both the fluctuation of $\psi$,
$
\Delta \psi = \langle (\psi - \langle \psi \rangle)^2 \rangle/\langle \psi \rangle
$\cite{siebertCriticalBehaviorActive2018},
which shows a peak at the transition point. Moreover, the boundary is also predicted by an unsupervised machine learning (ML) approach (dashed line in Figure~\ref{fig:fig1}(a)) \cite{xuRecognitionPolymerConfigurations2019} that does not rely on any empirical order parameter definitions (the details of these methods are discussed in Sec.~II of the SM \cite{supplemental1}). The phase boundaries predicted by the two methods are highly consistent, located at $(\Delta qE)_c=0.76$.

\begin{figure}[t]  
  \centering
  \includegraphics[width=0.48\textwidth]{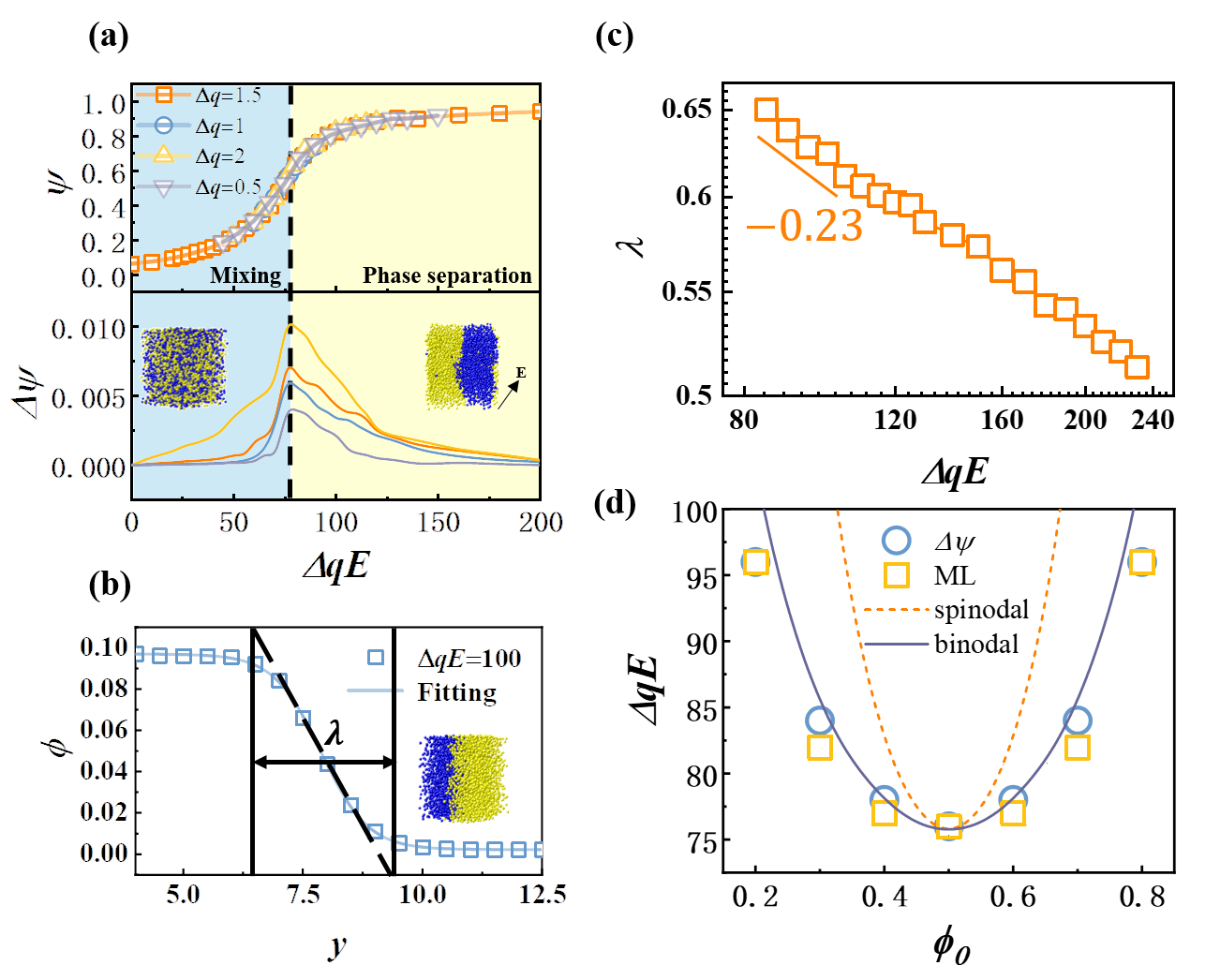}  
  \caption{Phase separation behavior in a binary system in 3D (a) $\psi$ and its fluctuation $\Delta \psi$ as functions of the driving force $\Delta q E$ for $\phi_0 = 0.5$. The phase-separated point is identified at $(\Delta q E)_c = 76$, via both unsupervised learning (vertical dashed line) and the peak position of $\Delta \psi$. (b) A typical snapshot of the interface between the A and B bulk phases. The interfacial width $\lambda$ can be determined based on mean-field theory fitting, e.g., $\Delta qE = 100$. (c) $\lambda$ follows a power law with the microscopic driving force, described by $\lambda \propto (\Delta q E)^{-0.23}$. (d) The phase diagram of laning in the 3D binary system was determined from simulation (blue circles) and unsupervised machine learning (yellow squares). The spinodal line (orange dashed line) and the binodal line (blue solid line) are predicted by mean-field theory.
  }
  \label{fig:fig1}
\end{figure}

This laning behavior can be described in a framework which reflects the balance between translational entropy and effective interactions, analogous to that of equilibrium phase separation to some extent. According to Onsager's variational principle, the characteristic functional of a laning system should be a Rayleighian instead of a free-energy functional \cite{SoftMatterPhysics2013}. 
Here, a time-scale separation assumption is adopted that the relaxation dynamics along $\mathbf E$ is on a faster timescale, while the lateral dynamics are relatively slower. This implies the decoupling of the longitudinal and lateral dynamical processes, and the convective flows are ignored in this framework. At the time scale of lateral phase separation, the longitudinal steady state is reached. Subsequently, small columnar clusters along $\mathbf E$ are formed with a drift velocity of 
$
v_{\alpha,z}  = q_{\alpha} E/\xi 
$.
Each cluster extends across the full length of the system in the z-direction and has a relatively larger lateral size compared to that of active particles. These clusters can be considered as pseudo-particles of component $\alpha$ which can diffuse laterally perpendicular to  $\mathbf E$ and the laning phenomenon is the lateral phase separation of these pseudo-particles.

According to the Onsager principle, the dynamic equations of these pseudo-particles can be determined by minimizing the Rayleighian 
$
\mathcal R = \Phi + \dot{A}.  
$
The free energy $A$ is solely contributed by the translational entropy of pseudo-particles in the lateral direction, 
$
A/L_z = \int \mathrm{d} \mathbf{R} \sum_{\alpha} \phi_{\alpha}\ln\phi_{\alpha} ,
$
where $\phi_{\alpha}(\mathbf R)$ is coarse-grained volume fraction of pseudo-particles.
The dissipation functional $\Phi$ is from the relative velocities of neighboring pseudo-particles, which can be decomposed into longitudinal contribution and lateral contribution, as
$
\Phi/L_z = \Phi_{v_z}/L_z + \Phi_{v_\mathbf{R}} /L_z.
$
The longitudinal term from the differences in drift velocities in the $z$ direction is the driving force of phase separation,
\begin{eqnarray} 
\Phi_{v_z}/L_z
= \int \mathrm{d} \mathbf{R}\left[ u_{AA}\phi_A^2 + 2u_{AB}\phi_A\phi_B +u_{BB}\phi_B^2 \right].
\end{eqnarray} 
Here, $u_{\alpha\beta}\equiv\frac{1}{2}\xi_z\left[{v}_{\alpha,z}(\mathbf {R}) - {v}_{\beta,z}(\mathbf {R}) \right]^2$ and $\xi_z$ are the longitudinal friction coefficient between pseudo-particles.
The change of longitudinal contribution from mixing is
\begin{eqnarray}\label{Phi_z}
 \Delta\Phi_{v_z}/L_z&=&\Phi_{v_z}/L_z-\Phi_{v_z,0}/L_z \nonumber\\
 &=& \tilde\chi \int \mathrm{d} \mathbf{R}\phi(1-\phi),
\end{eqnarray}
where
$
\Phi_{v_z,0}/L_z  
 = \int \mathrm{d} \mathbf{R}\left[ u_{AA}\phi_A +u_{BB}\phi_B \right] 
$
is the reference term before mixing and  
$
\tilde\chi\equiv  2u_{AB} -u_{AA}-u_{BB} $.
Considering the time-scale separation assumption, for a sufficiently long time period $\tau$, the longitudinal term $\Phi_{v_z}/L_z$ can be approximated as a time-invariant quantity, i.e., $\int_{\tau} {\rm d} t \Phi_{v_z}/L_z = \tau\Phi_{v_z}/L_z$. Formally, its contribution together with the lateral translational entropy can construct an effective \textit{free energy} $\tilde{A}$ 
, whose contribution to the Rayleighian can be written as 
\begin{eqnarray} 
 \dot{\tilde{A}}/L_z = \Delta\Phi_{v_z}/L_z + \dot{A}/L_z 
= \frac{\partial}{\partial t}(\Delta\mathcal{H} + A)/L_z, 
\end{eqnarray}
where
$
\Delta\mathcal{H}/L_z  = \chi \int \mathrm{d} \mathbf{R}\phi(1-\phi),
$
with an effective interaction parameter defined as
$
   \chi = \tau \tilde\chi.
$ This effective \textit{free energy} functional becomes 
\begin{equation}
   \tilde A/L_z = \int \mathrm{d}\mathbf{R} (\sum_\alpha \phi_\alpha \ln \phi_\alpha + \chi\phi_\alpha \phi_\beta )
\end{equation}
which formally coincides with the free energy governing phase separation in equilibrium binary systems. The phase separation arises from the competition between lateral translational entropy and the effective interaction, which stems from the relative difference in drift velocities. Furthermore,
by minimizing the Rayleighian functional 
$
\mathcal R/L_z = \Phi_{\mathbf{v}_{\mathbf{R}}}/L_z+\dot{\tilde{A}}/L_z,
$
one can obtain a Smoluchowski-type dynamic equation for the lateral phase separation of pseudo-particles. See Sec.~III of SM  \cite{supplemental1} for the theoretical picture of the laning based on the Onsager principle.

This picture manifests the formal similarity between the laning phenomenon and equilibrium phase separation. 
Based on this similarity, we propose that the interfacial properties of the laning system can draw on Landau's mean-field theory with
$F(\phi) = \int {\rm d} \mathbf{r}  [f_0 + c (\nabla \phi)^2]$ where $c$ is a constant.
The interfacial width $\lambda$, or equivalently the correlation length, can be determined by fitting the interface profile extracted from simulations to a hyperbolic tangent function, as shown in Figure~\ref{fig:fig1}(b) for $\Delta qE=100$. By varying $\Delta qE$ (See Sec.~V of the SM~\cite{supplemental1}), a relation between $\lambda$ and the microscopic driven force $\Delta qE$ can be obtained (Figure~\ref{fig:fig1}(c)). $\lambda$ decreases with increasing $\Delta qE$, following the scaling relation $\lambda\sim (\Delta qE)^{-k}$ with $k=0.23$. Two key parameters, both interfacial tension $\gamma$ and effective interaction parameter $\chi$  can be predicted according to mean-field theory by: (i) $\gamma = 2 \sqrt{2} c \phi_0^2/(3 \lambda)$ and (ii)$\lambda=(c/(2-\chi))^{1/2}$\cite{chaikin1995principles}. The interfacial tension is ready to be used to verify Young's relation in partial wetting.
Besides, with $\chi$ the phase diagram of phase separation consisting of the binodal line $\chi_b = \frac{1}{1-2\phi_0}\ln(\frac{1-\phi_0}{\phi_0})$, and the spinodal line $\chi_s(\phi) = \frac{1}{2} \left( \frac{1}{\phi} + \frac{1}{1 - \phi} \right)$ can be obtained \cite{SoftMatterPhysics2013,zhangPredictingFloryHugginsSimulations2017} which are compared to the phase boundaries from simulation determined by both $\Delta \psi$ and ML in Figure~\ref{fig:fig1}(d). The simulation results agree well with the binodal line, and the spinodal line delineates the unstable region within the phase-separated regime. This agreement, particularly for $\phi_0 \neq 0.5$, suggests that the proposed picture of phase separation of pseudo-particles provides a reasonable description of the laning phenomenon. It also indicates that the interfacial tension predicted using the same theory is credible.  

In a ternary system, the partial wetting phenomenon involves the aggregation of component C on the A-B interface. Considering $ q_A = 1 $, $ q_B = -1 $ and $ q_C = 0 $ as an example, the lateral distribution, 
$
\phi_{C}(\mathbf{R}) = \int \phi_{C}(\mathbf{r}) \, {\rm d}z
$ and typical snapshot as shown in Figure~\ref{fig:fig2}(a) and (b), respectively. 
For a sufficiently strong external field ($ E = 150 $), C particles form a C-rich thin film between the A and B bulk phases. Because $\chi_{AB} > \chi_{AC}=\chi_{BC}$, the A-B interface disappears, which leads to component C completely wetting the A-B interface. In the weak-field regime ($ E = 70 $), where both $\chi_{AC}$ and $\chi_{BC}$ are below the phase separation threshold, C particles are dispersed throughout the system, which is a drying state. 
For intermediate field strengths, $ E \in [95, 120] $, C particles accumulate at the interface by forming a lens-shaped structure in the cross section, as seen in the middle panel of Figure~\ref{fig:fig2}(a) and (b). It is a partial-wetting state, characterized by the emergence of three distinct interfaces: A-B, A-C, and B-C. 

\begin{figure}[t]  
  \centering
  \includegraphics[width=0.48\textwidth]{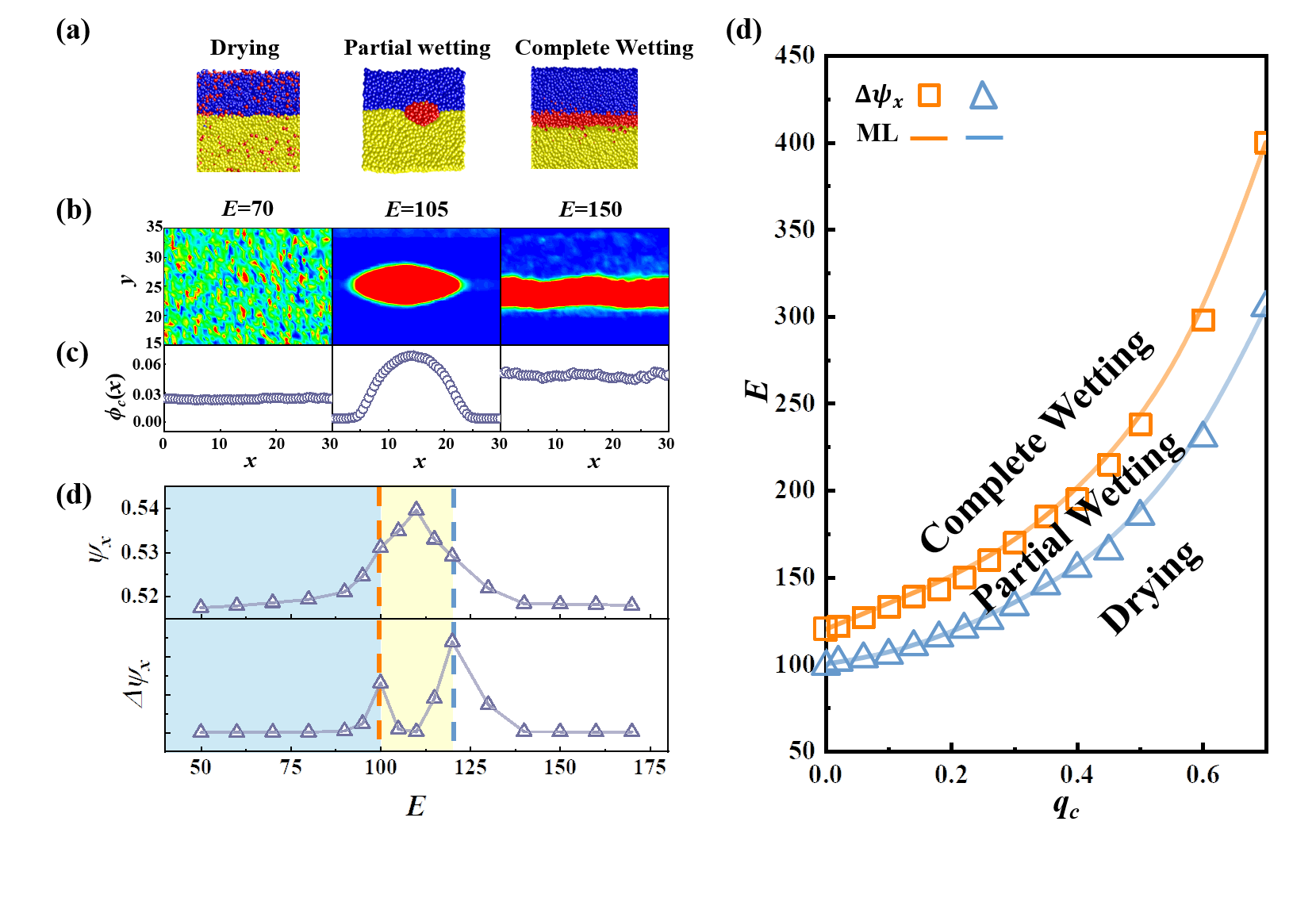}  
  \caption{Wetting phenomena in ternary laning system for $q_A=1$ and $q_B=-1$ (a) Typical snapshots of drying, partial wetting, and complete wetting states. Component C is colored red. (b) Lateral distribution of C particles, $ \phi_C(\mathbf{R}) $, for $ E = 70, 105 $ and $150 $, respectively.  (c) Aggregation of C particles along the interface, $ \phi_C(x) $. (d) $ \psi_x(x) $ and $ \Delta \psi_x $ as functions of $ E $. The orange and blue dashed lines indicate the transition points of drying-partial wetting, and partial wetting-complete wetting, respectively, as identified by ML. (e) Phase boundaries obtained via ML (yellow and blue curves) and that via $ \Delta \psi_x $ (symbols).}
  \label{fig:fig2}
\end{figure}

In order to clarify the emergence of partial wetting between drying and complete wetting, the aggregation of component C near the interface $
\phi_{C}(x) = \int_Y \phi_{C}(\mathbf{R}){\rm d}y
$ is shown in Figure~\ref{fig:fig2}(c)
where $Y\in [y_0-D,y_0+D]$, $y_0$ denote the A–B interface position and $D=10$. 
In both drying state ($ E = 70 $) and wetting state ($ E = 150 $), $ \phi_C(x) $ exhibits a homogeneous distribution along the interface. While it has a relatively high value in the wetting state, a C-rich film is formed. In the partial-wetting state ($ E = 105 $), C-rich film shrinks, resulting in a pronounced band in $ \phi_C(x) $, as shown in the middle panel of Figure~\ref{fig:fig2}(c). An order parameter
$
\psi_x = \frac{1}{n} \sum_{i=1}^{n} \left[\frac{n_{+}(i) - n_{-}(i)}{n_{+}(i) + n_{-}(i)}\right]^2
$   
can be used to describe these characters. 
The parameters $ n_{+}(i) $ and $ n_{-}(i) $ enumerate same-type and cross-type particles within an interval of $ 0.75\sigma $ centered on particle $i$ in the $x$-direction, where A and B are considered to be identical in type. The upper panel of Figure~\ref{fig:fig2}(d) presents $ \psi_x $ as a function of $ E $ for $ q_C = 0 $, exhibiting a peak when component c has the maximal aggregation along the interface. In both the drying and complete wetting states, $\phi_C(x)$ remains homogeneous, leading to $ \psi_x \to 0 $. The transition points between drying and partial wetting, as well as between partial wetting and complete wetting, can be determined by the fluctuation of $ \psi_x$,  
$
\Delta \psi_x = \langle (\psi_x - \langle \psi_x \rangle)^2 \rangle/\langle \psi_x \rangle
$.  
ML is also used to eliminate subjective bias in predicting phase boundaries.
$ \Delta \psi_x $ exhibits two distinct peaks at $ E = 100 $ and $ E = 125 $ (lower panel of Figure~\ref{fig:fig2}(d)), corresponding to the drying-to-partial wetting and partial wetting-to-wetting transitions, respectively, which are in excellent agreement with that from ML indicated by vertical dashed lines.
Based on these methods, the phase diagram of partial wetting in the ternary laning system for condition of $q_A=1$ and $q_B=-1$ is constructed (Figure~\ref{fig:fig2}(e)). 
In the drying state, the compatibility between C and A increases with the increase of $ q_C $, causing the C particles to preferentially distribute in the A-rich phase. And it needs a higher $ E $ to induce the phase separation of A and C. Consequently, both transition points move to higher values of $ E $ as $ q_C $ increases. Particularly, as $ q_C \rightarrow q_A $, the values of $ E $ for these transitions diverge.

In this nonequilibrium partial-wetting phenomenon, the dynamic properties (See Sec.~VI of the SM~\cite{supplemental1}) and the static structures of the C-rich phase are intriguing research topics that differ from those of the liquid droplet partial wetting on a gas-solid interface. Liquid droplets exhibit rotational symmetry about their central axis perpendicular to the gas-solid interface, and the contact line is a circle. In contrast, the C-rich phase observed here forms a columnar structure that is translationally invariant along the direction of $ \mathbf{E} $, and the contact lines are two straight lines. 
Typical simulation snapshots of the C-rich phase for $ q_C = 0 $ and $ 0.3 $ at $ E = 120 $ are compared in Figure~\ref{fig:fig3}(a). When $ q_C = 0 $, $\chi_{AC}=\chi_{BC}$ resulting in a symmetric lens-shaped cross section. In contrast, for $ q_C = 0.3 $,  $\chi_{AC}<\chi_{BC}$ leads to a larger C-A interface area compared to the C-B interface and an asymmetric lens-shaped cross section. The compatibility-induced change of the shape of the C-rich phase implies that the interfacial tension governs this phenomenon.
Referring to the definitions of the gas-liquid-solid system, the interfacial tensions and contact angles in the present system are defined in the cross section as illustrated in Figure~\ref{fig:fig3}(b)\cite{yuanDynamicsDissolutiveWetting2017}.

\begin{figure}[t]  
  \centering
  \includegraphics[width=0.48\textwidth]{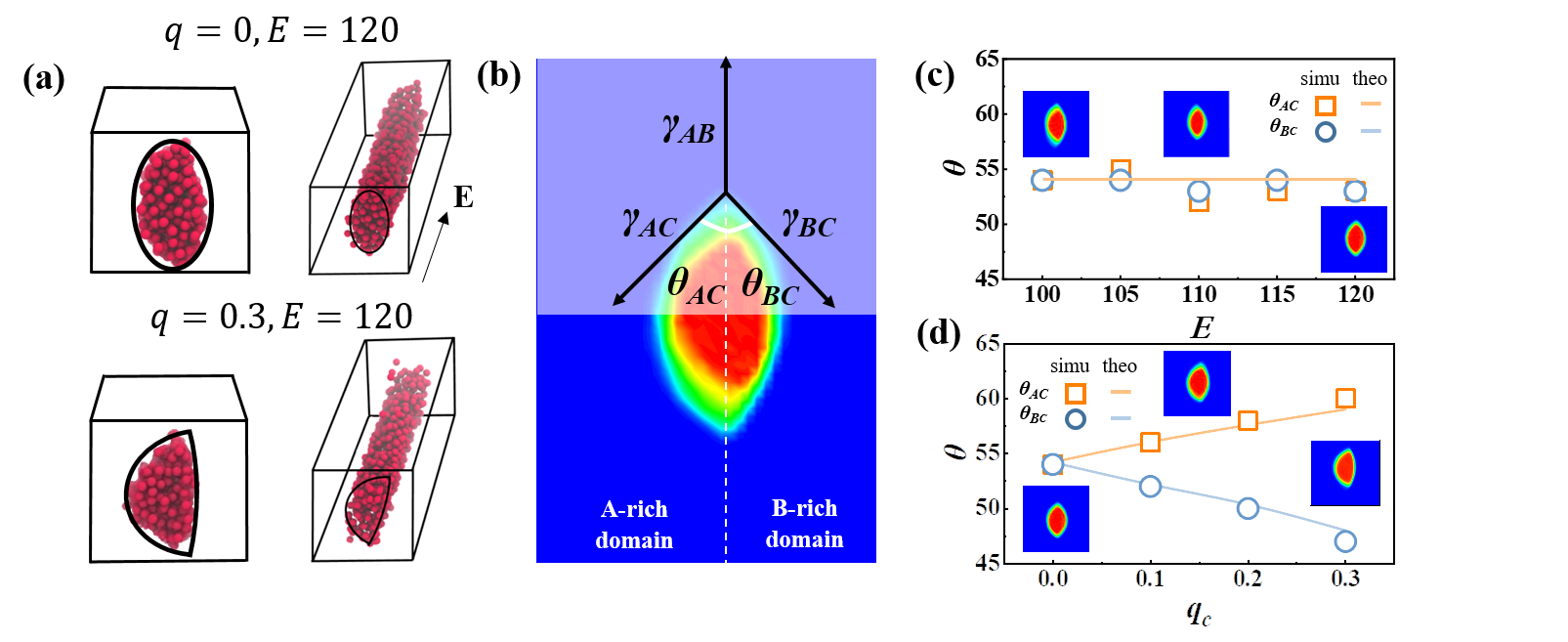}  
    \caption{(a) Typical Snapshots of the C-rich phase at $ E = 120 $, which have the shape of the column with symmetric ($ q_C = 0 $) and asymmetric ($ q_C = 0.3 $) lens-shaped lateral cross-section. (b) Definitions of contact angles and interfacial tensions. The quantitatively verification of Young's equation in partial wetting regime, (c) for $\theta_{\alpha, C}$ vs. $E$, with $ q_C = 0 $ as an example and (d) for $\theta_{\alpha, C}$ vs. $q_C$, with $ E = 120 $. The contact angles (symbols)  quantitatively match the predictions (curves) from Young's equation.} \label{fig:fig3}
\end{figure}

One major contribution of the present work is to quantitatively verify Young's equation in partial wetting of active matter. In parallel and perpendicular directions to the A-B interface, the equations are \cite{safranStatisticalThermodynamicsSurfaces2018,semprebonApparentContactAngle2017,jonosonoMolecularDynamicsStudy2024}:  
\begin{eqnarray}
\gamma_{AC} \cos(\theta_{AC}) + \gamma_{BC} \cos(\theta_{BC}) &=& \gamma_{AB} \nonumber
\\
\gamma_{AC} \sin(\theta_{AC}) - \gamma_{BC} \sin(\theta_{BC}) &=& 0.
\end{eqnarray}
According to the theoretical picture proposed above, the interfacial tension and microscopic driving force satisfy
$
\gamma_{\alpha\beta} \propto (\Delta q_{\alpha\beta} E)^{0.23}.
$  
Because $ \Delta q_{AB} = 2 $, $ \Delta q_{AC} = 1 - q_C $, and $ \Delta q_{BC} = 1 + q_C $, the contact angles satisfy:  
\begin{eqnarray}\label{Young_q}
(1 - q_C)^k \cos(\theta_{AC}) + (1 + q_C)^k \cos(\theta_{BC}) &=& 2^k, \nonumber \\
(1 - q_C)^k \sin(\theta_{AC}) -(1 + q_C)^k \sin(\theta_{BC})&=&0 .
\end{eqnarray}
These equations indicate the contact angles depend only on $ q_C $ and are independent with $E$ in partial wetting of laning system. 

The contact angles $ \theta_{AC} $ and $ \theta_{BC} $ obtained from simulations (orange and blue symbols) are compared with the predictions from the mean-field theory Eq.~(\ref{Young_q}) (solid curves) under different $ q_C $ and $ E $ in Figure~\ref{fig:fig3}.
When $ q_C = 0 $, in the partial wetting regime, $ E \in [100,125] $ (Figure~\ref{fig:fig3} (c)), the two contact angles are approximately equal and exhibit almost no dependence on $ E $. The mean-field theory predicts $ \theta_{AC} = \theta_{BC} = 54^\circ $ which is in excellent agreement with the simulation results. At the boundary between partial wetting and complete wetting, the contact angles abruptly drop to $ \theta_{AC} = \theta_{BC} = 0^\circ $, while at that between partial wetting and drying, they jump to $ \theta_{AC} = \theta_{BC} = 180^\circ $. This discontinuous transition is a distinctive feature of the present model, which makes the precise measurement of the contact angles near the phase boundary challenging. Therefore, to confirm the presence of these transitions and determine the phase boundaries, both the phenomenological order parameter $\psi_x$ and the ML methods were employed. Further examples for $q_C = 0.3$ and $q_C = 0.5$ are provided, which also verify that the contact angles are independent of $E$ (see Sec.~VII of the SM \cite{supplemental1}). Given $E$, for example $ E = 120 $, (Figure~\ref{fig:fig3}(d)), as $ q_C $ increases, the contact angles exhibit opposite trends: $ \theta_{AC} $ increases from $ 54^\circ $ to $ 60^\circ $, whereas $ \theta_{BC} $ decreases from $ 54^\circ $ to $ 46^\circ $. The mean-field theoretical predictions quantitatively match the simulation results. These findings demonstrate that nonequilibrium partial-wetting phenomena are also governed by the balance of interfacial tensions on contact lines, which described by Young's equation. 

This study clearly demonstrates partial wetting of active matter in a ternary laning system and verifies the shape of the C-rich phase governed by Young's equation. A phase diagram describing the transitions among complete wetting, partial wetting and drying are constructed. These results show that this phenomenon is analogous to that in the equilibrium system. A theoretical picture is proposed to explain this similarity, in which the longitudinal dissipation term from the differences in drift velocities gives rise to the effective interaction coefficient $\chi$ of phase separation.
This discovery extends the theoretical applicability of Young's equation to active systems. In terms of the structural design of nonequilibrium systems, dynamically manipulating interfacial tension by tuning the self-propelled  characteristics of active components (such as motility speed) offers a novel approach for developing intelligent, responsive active coatings, microfluidic devices and structured robotic swarms through interfacial wetting behavior.

\textit{Acknowledgments}—This work was supported by the National Natural Science Foundation of China (NSFC) No. 22173004, and the Fundamental Research Funds for the Central Universities No. 2024JBZX029.

\bibliography{ref}

\clearpage
\onecolumngrid
\appendix
\section*{Supplemental Material for \textit{“Partial-Wetting Phenomena in Active Matter”}}
\addcontentsline{toc}{section}{Supplementary Material}
\setcounter{figure}{0}
\setcounter{table}{0}
\setcounter{section}{0}
\setcounter{equation}{0}
\renewcommand{\thefigure}{S\arabic{figure}}
\renewcommand{\thetable}{S\arabic{table}}
\renewcommand{\thesection}{S\arabic{section}}
\renewcommand{\theequation}{S\arabic{equation}}
\section{1.Model and method}
\subsection{1.1.Model}
To describe the wetting behavior of active particles in a nonequilibrium system, this study establishes a three-dimensional laning system model to analyze the phase separation points and interfacial properties between coexisting phases to describe the wetting behavior of active particles in a nonequilibrium system. In a three-dimensional cubic box with volume $ V$, the system consists of three types of active particles A, B, and C, with a total particle number $ n = \sum_{\alpha} n_\alpha$, where $\alpha = A, B, C$. The average particle number density of the system is set as $\bar{\rho} = n/V = 0.5$. Under the external field $\mathbf{E}$, the force acting on the $ i $-th particle is $\mathbf{f}_{i,\alpha} = \mathbf{E} q_\alpha$, where $ q_\alpha \in [-1,1] $ represents the coupling coefficient between particle type $\alpha$ and the external field, which can also be interpreted as the particle "charge". Its magnitude controls the strength of the force exerted on the particle, and its sign determines the direction of the force. By adjusting the external field coupling coefficient, a laning transition can be induced in the system, in which active particles form a stable phase separation pattern near the interface.

The state of the system is represented by the positions of all particles $\{ \mathbf{r}_i\}_{i=1}^{N} $. Their dynamics are governed by coupled overdamped Langevin equations:
\begin{eqnarray}
\partial_t \mathbf{r}_i = \frac{D_t}{k_B T} \left(\mathbf{F}_{\text{ex}}(\{\mathbf{r}\}) + \mathbf f_{i,\alpha} \right) + \sqrt{2D_t} \mathbf{\eta}_i^T,
\end{eqnarray}
where $k_B$ is the Boltzmann constant, $T$ is the absolute temperature, and $D_t$ is the translational diffusion constant. $\mathbf{F}_{\text{ex}}$ arises from the excluded volume interactions between particles, which in this study are described by the Weeks–Chandler–Andersen (WCA) potential, $V_{\text{ex}} = 4\epsilon[(\frac{\sigma}{r})^{12} - (\frac{\sigma}{r})^6] + \epsilon$ \cite{rednerStructureDynamicsPhaseSeparating2013b}, where $r$ is the distance between the centers of two particles and $\sigma$ is the particle diameter. The parameter $\epsilon$ ensures the continuity of the potential at the cutoff distance $r = 2^{1/6} \sigma$. $\eta$ is a Gaussian white noise variable, characterized by $\langle \eta_i(t) \rangle = 0$ and $\langle \eta_i(t) \eta_j(t') \rangle = \delta_{ij} \delta(t - t')$.
To nondimensionalize the equation of motion, we choose the particle diameter $\sigma$ as the unit of length, the thermal energy $k_B T$ as the unit of energy, and the characteristic time $\tau = \sigma^2/D_t$ as the unit of time. Molecular dynamics methods are employed for numerical simulations, with the maximum time step set to $1 \times 10^{-4} \tau$. The properties of the system depend on two parameters: the mixing ratio, characterized here by the average volume fraction of component A ($\phi_0$), and the relative external force acting on different components ($f_{i,\alpha}$). Conventional laning studies are typically conducted in two-dimensional space, where the A-B interface in the direction perpendicular to the external field reduces to a point. However, this study focuses on the wetting phenomenon, which involves the inhomogeneous distribution of the third type of particle along the interface. This feature cannot be captured in two-dimensional systems. Therefore, the present work is carried out in three-dimensional space, where the A-B interface becomes a line in the plane perpendicular to the external field. The wetting behavior can then be studied through the inhomogeneous distribution of component C along this line. The direction of the external field is defined as the $z$-direction, and the vector perpendicular to the external field is denoted as $\mathbf{R}$. The normal direction of the A-B interface is along the $y$-direction, while the direction parallel to the interface is along the $x$-direction.
The numerical update algorithm of the system is as follows:
\begin{eqnarray}
 v_{\alpha=x,y} = a F_{\text{ex}}(\mathbf{r_i}) + b \eta\\
 v_{\alpha=z} = a \left( F_{\text{ex}}(\mathbf{r_i}) + f_{i,\alpha} \right) + b \eta\\
 \mathbf{r} = \mathbf{r} + \mathbf{v}
\end{eqnarray}
Specifically, the correlation coefficient is defined as:
\begin{eqnarray}
a = \frac{D_t}{k_B T}, \quad b = \sqrt{2 D_t}
\end{eqnarray}

\subsection{1.2.Simulation of the binary system}
The core objective of this study is to quantitatively verify that the partial-wetting phenomenon in active systems satisfies Young’s equation, which requires establishing the relationship between the interfacial tension of the interface between two bulk phases and the microscopic driving force. Therefore, in a binary active system composed of particles A and B, the conditions for phase separation are determined, and the width of the interface formed between the coexisting A and B phases is measured. First, the phase separation point of the binary system is identified. A simulation system is constructed in a three-dimensional cubic box with dimensions $L_x = L_y = L_z = 20$, and periodic boundary conditions are applied in all three directions. The total number of particles is set to $n = 4000$. The initial state of the system is a randomly mixed configuration, and the simulation is run under a given external field for a sufficiently long time $t = 20 \tau$ to ensure a steady state is reached. The order parameter is then measured in the time interval from $t = 20 \tau$ to $t = 100 \tau$.

To measure the interfacial width, the initial configuration is set such that one half of the simulation box is occupied by pure component A, and the other half by pure component B. An interface between the A and B phases is formed through mutual diffusion of A and B particles during the simulation. After the system reaches a steady state at $t = 20 \tau$, the particle distribution in the steady state is analyzed to obtain a stable interfacial profile. Based on $\phi(y)=\int\phi(\mathbf{r})\, {\rm d}x\, {\rm d}z$, where $\phi(\mathbf{r})=\langle\sum_{i=1}^{n_A}\delta(\mathbf{r}_{i}^{\alpha=A}-\mathbf r)\rangle$ denotes the volume fraction of particle A at position $\mathbf{r}$, we further apply a fitting method to determine the interfacial width (see Sec.5 for details).

\subsection{1.3.Simulation of the ternary system}
To study the behavior of partial wetting, under conditions where components A and B undergo phase separation and form a stable A-B two-phase interface, a third type of active particle C is introduced into the interfacial region to construct a ternary active system. The initial configuration of the system is a sandwich structure of A–B–C, where component C is initially positioned near the A-B interface. The simulation space is a cubic box with dimensions $L_x = L_y = L_z = 40$, and periodic boundary conditions are applied in all three directions. The total number of particles is set to $n = 13{,}500$, with the fraction of components A, B, and C given by $\phi_A = 0.56$, $\phi_B = 0.38$, and $\phi_C = 0.06$, respectively. The system reaches a steady state at simulation time $t = 200\tau$, and remains stable throughout the total evolution time of $1000\tau$. All measurements of order parameters are performed in the steady-state interval $t \in [200\tau, 1000\tau]$. The coupling coefficients between components A and B and the external field are fixed at $q_A = 1$ and $q_B = -1$, respectively. Under strong driving conditions $E > 50$, the condition $\Delta q_{AB} E > (\Delta qE)_c$ is satisfied, ensuring that components A and B undergo clear phase separation, forming two bulk regions distinctly separated by an interface. The coupling strength of component C, $q_C$, is varied within the interval $q_C \in [0,1]$, allowing for continuous tuning of the drift velocity, and thereby altering the effective interaction parameter $\chi_{\alpha\beta}$ between pseudo-particles. When $q_C = 1$, $\chi_{AC} = 0$, indicating complete miscibility between components A and C, while $\chi_{BC}$ reaches its maximum; conversely, when $q_C = 0$, $\chi_{AC} = \chi_{BC}$, meaning that component C interacts equally with both components A and B.
\newpage

\section{2.Determining phase transition boundaries} 
\subsection{2.1.Determining phase transition boundaries using order parameters in binary system} 
To determine the phase separation point in a binary system, we fix the initial volume fraction of component A at  $ \phi_0 = 0.5 $, and set the coupling coefficient of A-type particles with the external field to $ q_A = 1 $. The coupling coefficient of B-type particles $ q_B $ is varied to control $ \Delta q = q_A - q_B $, where $ q_B $ takes values of 0.5, 0, -0.5, and -1, corresponding to $ \Delta q $ values of 0.5, 1, 1.5, and 2, respectively. For each $ \Delta q $, the external field strength $ E $ is varied such that $ \Delta q E $ ranges from 0 to 200, and the order parameter $ \psi $ is measured to characterize the degree of phase separation in the system. As shown in the upper part of Fig. 1(a) in the main text, with increasing $ \Delta qE$, the system transitions from a mixed state to a phase-separated state. The curves for different $ \Delta q $  overlap, indicating that the parameter $ \Delta q E $ serves as the driving force for phase separation.
To quantitatively determine the phase separation point, we compute the fluctuation of the order parameter $ \Delta \psi $, with the results shown in the lower part of Fig. 1(a) in the main text. The peak position of $ \Delta \psi $ corresponds to the phase separation point. It is observed that the phase separation points for systems with different $ \Delta q $ coincide. However, the absolute values of $ \Delta \psi $ near the phase separation point vary across different $ \Delta q $ values.

At steady state, the drift velocity of individual particles depends on $ q_\alpha E $. In binary systems, an appropriate reference frame can always be chosen to ensure that the mean velocity of the system is zero. Consequently, the phase behavior and correlation length depend only on the parameter $ \Delta qE $. However, near the phase separation point, the fluctuations of the order parameter $ \Delta \psi $ are not solely determined by $ \Delta q E $. As shown in Figure 1(a) in the main text, $ \Delta \psi $ varies for different values of $ \Delta q $ near the critical point. For systems with small coupling coefficient differences, such as $ \Delta q = 0.5 $, larger values of $ E $ lead to smaller fluctuations. Conversely, for systems with large coupling coefficient differences, such as $ \Delta q = 1.5 $, smaller values of $ E $ result in larger fluctuations. This behavior arises because the external field strength $ E $ determines the force acting on individual particles, which in turn sets the persistence length of their drift motion. A longer persistence length suppresses fluctuations, leading to a more stable phase-separated state. Representative particle configurations are also shown in the figure: the left panel displays the mixed state, and the right panel shows the structure after phase separation, where stripes align along the field direction, reflecting the formation of ordered structures induced by the drive.

In this study, to verify the applicability of the mean-field theory used to predict surface tension based on interfacial width, we also predict the relationship between the effective interaction parameter $\chi$ and the interface width using this theory. With this parameter, the complete phase diagram of the system, namely the phase separation points at different composition ratios $\phi_0$, can be theoretically predicted. Comparing theoretically predicted and simulation-obtained phase boundaries provides evidence for the applicability of the mean-field theory for the interface. The definitions of the order parameter $ \psi $ and its fluctuation $ \Delta \psi $ also identify the critical coupling strength for phase separation as $ (\Delta q E)_c \approx 76 $ under the condition $ \phi_0 = 0.5 $. Furthermore, we vary the volume fraction $ \phi_0 $ of component A from 0.1 to 0.9 and repeat the above analysis to obtain phase separation points under different component ratios, as shown by the blue points in Fig. 1(d) of the main text.

\subsection{2.2.Determining phase transition boundaries using unsupervised machine learning methods in binary system} 
Traditional approaches to studying phase separation typically rely on the definition of an order parameter. Although this method is intuitive, the definition of the order parameter often depends on prior knowledge of the physical properties of the system. This implies that for new and complex systems, it may be difficult to find a suitable order parameter to determine whether a phase transition has indeed occurred. In addition, the order parameter method is often not sufficiently precise in identifying the phase transition point, especially when the transition process is relatively continuous or involves multiple phase transition stages. In such cases, the change in the order parameter may not be significant enough to accurately capture the phase transition point. Based on these limitations, this study adopts an unsupervised machine learning (ML) approach to improve the accuracy and reliability of phase transition point identification.
Here, Principal Component Analysis (PCA), an unsupervised machine learning method, is used to analyze the configurational data of active particles. By diagonalizing the covariance matrix of the data, a set of mutually orthogonal normalized vectors (principal components) is obtained. These vectors are ordered by the magnitude of their eigenvalues and reflect the most significant modes of variation in the data as the driving force $ \Delta q E $ changes, namely the soft modes.

In the binary active system, this study constructs a dataset composed of observations in a three-dimensional parameter space with $\phi_0=0.5$ and 4000 particles. The dataset covers 100 time steps under steady state conditions for 16 different values of the external field strength $E$, ensuring that the data span across the phase transition point. By subtracting the mean from the original data matrix, a zero-mean data matrix $\mathbf{S}_{N \times D}$ is obtained, where $N$ is the number of samples and $D$ is the feature dimension. Then, the covariance matrix $\mathbf{C} = \mathbf{S}^T \mathbf{S}$ is computed, and eigenvalue decomposition is performed to express it as $\mathbf{S}^T\mathbf{S} W_l = \lambda_l W_l$($l = 1, 2, \dots, D$), where $\lambda_l$ are the eigenvalues and $W_l$ are the corresponding orthogonal eigenvectors. The eigenvalues are sorted in descending order,  $\lambda_1 \geq \lambda_2 \geq \lambda_3 \geq \dots \geq \lambda_D \geq 0$. The first $K$ eigenvectors with the largest eigenvalues are retained, and the low-dimensional representation of the configurational data is obtained through the projection transformation $Z_{N \times K} = X_{N \times D} W_{D \times K}$. The projection values can be considered as the order parameters obtained through unsupervised learning.

Figure~\ref{fig:fig1}(a) shows the eigenvalue distribution $\lambda_k$ of the first 20 principal components under the condition of volume fraction $\phi_0 = 0.5$ and $\Delta q_{AB} = 2$. It can be seen that the first eigenvalue $\lambda_1$ is significantly larger than the others, indicating that the main variation in the system is dominated by the first principal component. Furthermore, Figure~\ref{fig:fig1}(b) shows the distribution of the projection of particle coordinates onto the first principal component direction (the principal component scores) $W_1$ under different external field strengths $\Delta qE$. The results show that when $\Delta qE \approx 76$, the distribution of $W_1$ exhibits a significant jump, indicating that the system undergoes a transition from a uniformly mixed state to a phase-separated state. This behavior is further verified in Figure~\ref{fig:fig1}(c), which presents the variation of the probability distribution function $G(W_1)$ under different values of $\Delta qE$. The probability distribution evolves from a single-peaked to a double-peaked structure, reflecting the emergence of distinct phase regions in the system, which marks the occurrence of phase separation. The phase separation point predicted by ML is marked with a vertical dashed line in Figure 1(a) of the main text.

Similarly, the configurational data obtained from simulations under different composition ratios ($\phi_0 = 0.1, 0.2, \dots, 0.9$) are used as input, and ML is applied to identify the phase separation points under each ratio, represented by yellow square symbols in Figure 1(d) of the main text.

\begin{figure}[htbp] 
  \centering 
  \includegraphics[width=0.8\textwidth]{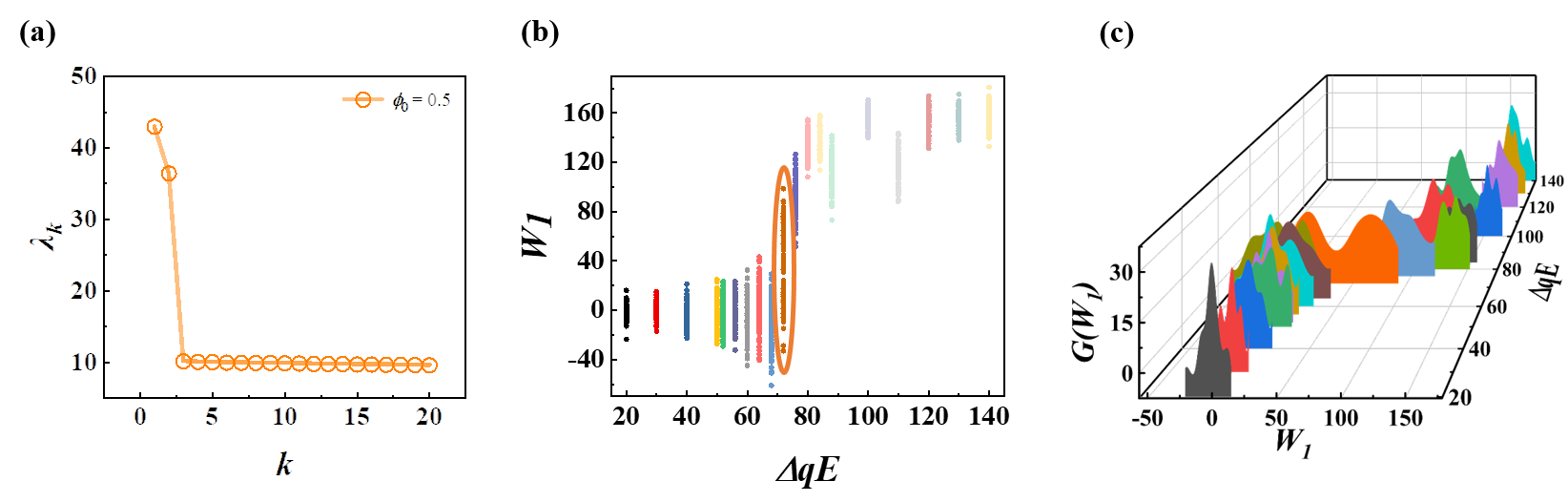} 
  \caption{Phase separation of binary active systems identified using unsupervised machine learning (ML) methods. (a) Distribution of eigenvalues $\lambda_k$ for the first 20 principal components. (b) Distribution of the projection values $W_1$ of each data point along the first principal component direction as a function of the parameter $\Delta qE$. (c) Probability distribution $G(w_1)$ of the projection values along the first principal component direction as a function of the parameter $\Delta qE$.}
  \label{fig:fig1}
\end{figure}

\subsection{2.3.Order parameters and unsupervised machine learning define phase boundaries in ternary systems} 
The core objective of this study is to determine whether the transitions between complete wetting–partial wetting and partial wetting–drying occur in a ternary system, and to identify the phase boundaries of these two transitions. To investigate the phase diagram of wetting behavior in the ternary system, we first fix the coupling coefficient of the C-type particles as $q_C = 0$, and under this condition, vary the external field strength $E$, while measuring the order parameter $\psi_x$ and its fluctuation $\Delta \psi_x$. As shown in Figure 2(d) of the main text, $E$ is gradually varied in the range $50 \leq E \leq 170$, and with increasing field strength, the order parameter $\psi_x$ exhibits a pronounced peak structure, indicating that the system undergoes a transition from a drying state to a partial-wetting state, and eventually to a complete wetting state. Furthermore, $\Delta \psi_x$ shows prominent peaks near $E = 100$ and $E = 125$, reflecting two wetting transitions at these field strengths. The results indicate that $E = 100$ and $E = 125$ correspond to the critical points of the drying–partial wetting and partial wetting–complete wetting transitions, respectively. On this basis, we further vary the coupling strength $q_C$ of the C-type particles with the external field in the range $0 \leq q_C \leq 0.7$, and repeat the above analysis for each $q_C$ value to obtain the phase diagram of wetting transitions, as shown in Figure 2(e) of the main text. In this diagram, blue triangles and orange squares denote the phase boundaries of the partial wetting–drying and complete wetting–partial wetting transitions, respectively.

To overcome the subjectivity in identifying the two transition points, we also employ unsupervised machine learning (ML) methods to process the particle configuration data. In the study of the ternary active system, this work focuses on the steady state behavior of the system under different external field coupling strengths of the C component ($q_C = 0, 0.1, 0.2, \dots, 0.7$). The constructed dataset contains the spatial information of 3500 C particles, covering the three-dimensional parameter space. Specifically, for each $q_C$ value, the data includes 12 different external field strengths $E$, and for each state, 200 time steps of particle coordinates ($X, Y$) are recorded at steady state, ensuring that the data spans the wetting transition region. Under steady-state conditions, only the lateral phase separation of the C component needs to be considered, so the two-dimensional coordinates ($X, Y$) of the C component can effectively capture the evolutionary characteristics of wetting behavior. The eigenvalue spectrum shown in Figure~\ref{fig:fig2}(a) indicates that the first principal component is significantly dominant, suggesting that PCA can effectively extract the main mode of variation in the system. When $\Delta q_{AC} = 1$, the projection of particle coordinates onto the first principal component direction $W_1$ as a function of external field strength $E$ is shown in Figure~\ref{fig:fig2}(b). The results reveal that the system undergoes two phase transitions around $E = 100$ and $E = 120$, corresponding to transitions from drying to partial wetting and from partial wetting to complete wetting, respectively. Furthermore, Figure~\ref{fig:fig2}(c) presents the probability distribution function $W_1$ of the principal component projection $G(W_1)$ as a function of $E$, clearly reflecting the dynamical evolution of the wetting state and providing an effective quantitative analysis method for understanding wetting behavior in nonequilibrium active systems.
\begin{figure}[htbp] 
  \centering 
  \includegraphics[width=0.8\textwidth]{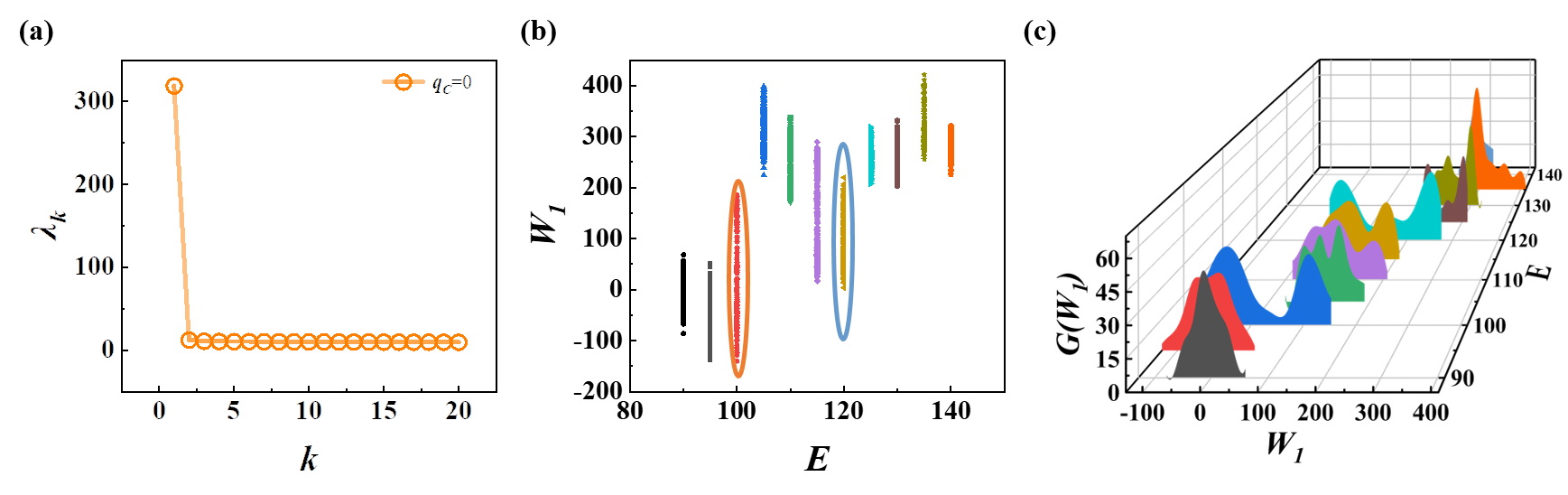} 
  \caption{Unsupervised machine learning (ML) method for constructing the wetting phase diagram in the three-dimensional ternary system. (a) Distribution of the eigenvalues $\lambda_k$ of the first 20 principal components. (b) Distribution of the projection values $W_1$ of each data point along the first principal component as a function of the parameter $E$. (c) Distribution of the probability density function $G(W_1)$ of the first principal component as a function of the parameter $E$.
  }
  \label{fig:fig2} 
\end{figure}
\newpage

\section{3.A functional for laning transition based on the Onsager principle theory} 
 Considering 2 types of particles A and B in a system with volume of $V$, with numbers of $n_A$ and $n_B$, respectively. The total number of particles is $n = n_A + n_B$. Under an external field $\bf{E}$, the forces acting on particle $\alpha = A, B$ is expressed by $\bf{f}_\alpha=\bf{E}q_\alpha$. Here
$q_{\alpha}$ is the "charge" of particle $\alpha$. The direction of the field $\bf{E}$ is defined as the z-axis direction. $\mathbf R$ is the vector in the x-y plane. In this section, $q_A=-q_B=1$ is considered, i.e., $\mathbf f_A=-\mathbf f_B$.

For the sake of finding a functional to describe the lateral phase separation of the laning system, it is assumed that $\mathbf{R}$ is a slow variable and $z$ is a fast variable. Namely, when the strength of the field is strong enough, a steady current forms in small clusters along the $z$ direction, with a velocity of 
\begin{eqnarray} \label{vz}
    v_{\alpha,z}  =\frac{q_{\alpha} E}{\xi}. 
\end{eqnarray}
The clusters mainly consist of the same type of particles to avoid collisions between different types of particles. The convective flow is ignored here.
 
Each cluster with a steady state can be considered as a pseudo-particle of $\alpha$ which can diffuse within a 2-dimensional space perpendicular to the z-axis. Because $\mathbf R$ is a slow variable, a coarse-grain length scale is considered in this direction, which is sufficiently large compared to the size of drift-diffusion particles $A$ and $B$ and sufficiently small to describe the lateral diffusion of the pseudo-particles. On this length scale, there are $n_\alpha$ clusters composed of $\alpha$ particles in the system. A volume fraction of $\alpha$ pseudo-particle in the 2-d space can be defined as
\begin{eqnarray}
    \rho_0\phi_\alpha(\mathbf R) = \sum_{i=1}^{n_\alpha}\delta\left(\mathbf{R}_{i\alpha}-\mathbf{R}\right).
\end{eqnarray}
Here, $n_\alpha$ is the number of pseudo-particle $\alpha$, and $\mathbf{R}_{i\alpha}$ is the position of the i-th pseudo-particle $\alpha$. The incompressible condition $\phi_A+\phi_B=1$ should be satisfied. Define $\phi = \phi_A =1-\phi_B$

The diffusion velocity of the pseudo-particle $\mathbf{v}_\alpha$ is decomposed to the velocity along the direction of the external field, which is only determined by Eq.~(\ref{vz}), and the lateral direction $\mathbf{v}_{\alpha,\mathbf R}$, which will be determined by  minimizing the Rayleighian 
$
\mathcal R = \Phi + \dot{A}.    
$
The free energy of a system consisting of pseudo-particles includes only the contribution from the translational entropy of the pseudo-particles in a two-dimensional lateral space, 
\begin{equation}\label{entropy}
A/L_z = \int \mathrm{d} \mathbf{R} \left[ \phi\ln\phi + (1+\phi)\ln(1+\phi) \right].
\end{equation}
Which favors the mixing of pseudo-particles. 

The dissipation function of the system $\Phi$ is from the relative velocity from neighboring pseudo-particles. It can be decomposed into longitudinal contribution along the direction of the external field and lateral contribution perpendicular to it, as
\begin{eqnarray} \label{dissipation}
\Phi/L_z 
&=& \frac{1}{2}\bar\xi_{\mathbf{R}} \sum_ \mathbf{R}\sum_{\mathbf{R}^\prime\in {NR}}\left[\mathbf{v}_{\alpha,\mathbf{R}} (\mathbf{R}) - \mathbf{v}_{\beta,\mathbf{R}} (\mathbf{R}^\prime)\right]^2 + 
    \frac{1}{2}\xi_{z} \sum_ \mathbf{R}\sum_{\mathbf{R}^\prime\in {NR}}\left[v_{\alpha,z} (\mathbf{R}) - v_{\beta,z} (\mathbf{R}^\prime)\right]^2\nonumber\\
         &=& \Phi_{v_\mathbf{R}} /L_z + \Phi_{v_z}/L_z.
\end{eqnarray}   
Here, the translational invariance along the direction of the external field and the dissipation function per unit length are considered. 
The first term is the dissipation from lateral diffusion of pseudo-particle, and $v_{\alpha,z} (\mathbf{R})$ is the lateral velocity of pseudo-particle located at $\mathbf{R}$. Consider a coarse-grain length scale in the lateral direction, which is much larger than that of the cross-section of the pseudo-particle. Then  
\begin{eqnarray} \label{Phi_R}
\Phi_{v_\mathbf{R}} /L_z
&=& \frac{1}{2}\bar\xi_{\mathbf{R}} \int \rm{d} \mathbf{R} \left[\mathbf{v}_{\alpha,\mathbf{R}} (\mathbf{R}) - \mathbf{v}_{\beta,\mathbf{R}} (\mathbf{R})\right]^2 .
\end{eqnarray}  
The volume conservation condition indicates that, $\mathbf{v}_{A,\mathbf{R}}$ and $\mathbf{v}_{B,\mathbf{R}}$ are not independent, but satisfy the relation
\begin{eqnarray} 
\mathbf{v}_{B,\mathbf{R}}
= -\phi/(1-\phi)\mathbf{v}_{A,\mathbf{R}}.
\end{eqnarray} 
Therefore, the dissipation term can be written as 
\begin{eqnarray}  
\Phi_{v_\mathbf{R}} /L_z
= \frac{1}{2}\xi_{\mathbf{R}} \int \rm{d} \mathbf{R} \left[\mathbf{v}_{A,\mathbf{R}} (\mathbf{R}) - \bar{\mathbf{v}}_{\mathbf{R}} (\mathbf{R})\right]^2 ,
\end{eqnarray} 
where the average velocity is
\begin{eqnarray}  
\bar{\mathbf{v}}_{\mathbf{R}}
= \phi\mathbf{v}_{A,\mathbf{R}} +(1-\phi)\mathbf{v}_{B,\mathbf{R}},
\end{eqnarray} 
and 
$
\xi_{\mathbf{R}} =  \bar{\xi}_{\mathbf{R}}/(1-\phi)^2.
$
In present work, $\bar{\mathbf{v}}_{\mathbf{R}}
=0$. Then
\begin{eqnarray}  
\Phi_{v_\mathbf{R}} /L_z
= \frac{1}{2}\xi_\mathbf{R} \int \rm{d} \mathbf{R} \mathbf{v}_{A,\mathbf{R}} (\mathbf{R})^2 .
\end{eqnarray} 

The last term of Eq.~(\ref{dissipation}) is the dissipation from relative motion between neighboring pseudo-particles along the external field, which is critical to the effective interaction for pseudo-particles, which drives the lateral phase separation of pseudo-particles.
It can be expressed as  
\begin{eqnarray} 
\Phi_{v_z}/L_z&=&  
    \frac{1}{2}\xi_z \sum_ \mathbf{R}\sum_{\mathbf{R}^\prime\in {NR}}\left[v_{\alpha,z} (\mathbf{R}) - v_{\beta,z} (\mathbf{R}^\prime)\right]^2\nonumber\\
         &=& \sum_{\alpha,\beta}\int \mathrm{d} \mathbf{R}\phi_\alpha(\mathbf {R})\frac{1}{2}\xi_z\left[{v}_{\alpha,z}(\mathbf {R}) - {v}_{\beta,z}(\mathbf {R}) \right]^2 \phi_\beta(\mathbf {R})\nonumber\\
 &=& \int \mathrm{d} \mathbf{R}\left[ u_{AA}\phi_A^2 + 2u_{AB}\phi_A\phi_B +u_{BB}\phi_B^2 \right].
\end{eqnarray} 
Here, $u_{\alpha\beta}\equiv\frac{1}{2}\xi_z\left[{v}_{\alpha,z}(\mathbf {R}) - {v}_{\beta,z}(\mathbf {R}) \right]^2$. 
Define
\begin{eqnarray}
    \Phi_{v_z,0}/L_z  
 = \int \mathrm{d} \mathbf{R}\left[ u_{AA}\phi_A +u_{BB}\phi_B \right] 
\end{eqnarray}
as the reference dissipation before mixing when the same component surrounds any component.
Then the change of dissipation from mixing is
\begin{eqnarray}\label{Phi_z}
 \Delta\Phi_{v_z}/L_z&=&\Phi_{v_z}/L_z-\Phi_{v_z,0}/L_z \nonumber\\
 &=& \int \mathrm{d} \mathbf{R}\left[ u_{AA}\phi_A^2 + 2u_{AB}\phi_A\phi_B +u_{BB}\phi_B^2  -  u_{AA}\phi_A -u_{BB}\phi_B \right] \nonumber\\
 &=& \int \mathrm{d} \mathbf{R}\left[ u_{AA}\phi_A(\phi_A-1) + 2u_{AB}\phi_A\phi_B +u_{BB}\phi_B(\phi_B-1)   \right]\nonumber\\
 &=& \int \mathrm{d} \mathbf{R}\left[  2u_{AB} -u_{AA} -u_{BB} \right]\phi_A\phi_B\nonumber\\
 &=& \tilde\chi \int \mathrm{d} \mathbf{R}\phi(1-\phi).
\end{eqnarray}
Here, 
\begin{eqnarray}
\tilde\chi&\equiv& 2u_{AB} -u_{AA}-u_{BB} \nonumber\\
&=&  \frac{\xi_z}{2}\left[2({v}_{A,z} - {v}_{B,z})^2-({v}_{A,z} - {v}_{A,z})^2-({v}_{B,z} - {v}_{B,z})^2\right]\nonumber\\
&=& \xi_z({v}_{A,z} - {v}_{B,z})^2. 
\end{eqnarray}
It should be noted that phase separation in the lateral direction is a slow variable. Its intrinsic time scale is slower than that of relaxing to the steady state of forming pseudo-particles. The time unit $\tau$ chosen here is sufficiently long, and then the longitudinal dissipative term of $\Phi_{v_z}/L_z$ is homogeneous. 
\begin{eqnarray}
\int {\rm d} t \Phi_{v_z}/L_z = \tau\Phi_{v_z}/L_z = \tau \tilde\chi \int \mathrm{d} \mathbf{R}\phi(1-\phi).
\end{eqnarray}
This contribution can be written as an effective interaction 
\begin{eqnarray}\label{chi_eff}
    \Delta\mathcal{H}/L_z \equiv 
    \chi \int \mathrm{d} \mathbf{R}\phi(1-\phi),
\end{eqnarray}
with the effective interaction parameter defined as
$
   \chi = \tau \tilde\chi.
$
Formally, this contribution together with the translational entropy Eq.~(\ref{entropy}), can construct an effective free energy $\tilde{A}$ 
whose contribution to the Rayleighian can be written as 
\begin{eqnarray} 
 \dot{\tilde{A}}/L_z = \Phi_{v_z}/L_z + \dot{A}/L_z 
= \frac{\partial}{\partial t}(\Delta\mathcal{H} + A)/L_z 
\end{eqnarray}
where
\begin{eqnarray}
    \tilde A/L_z &\equiv& \int \mathrm{d}\mathbf{R}\left[ \phi \ln \phi + (1-\phi) \ln (1-\phi)  + \chi\phi(1-\phi) \right ]  \nonumber \\
    &=&  \int \mathrm{d}\mathbf{R} \tilde {\mathcal{A}}[\phi]
.   
\end{eqnarray}
Here, effective free energy density $\tilde {\mathcal{A}}[\phi]$ is defined.
It is analogous to the free energy formation for equilibrium phase separation, 
$ 
f_{F-H} = \phi \ln \phi + (1-\phi) \ln (1-\phi)  + \chi\phi(1-\phi).    
$

Using \begin{equation}\label{conservation}
\frac{\partial \phi}{\partial t} =- \frac{\partial (\phi \mathbf{v}_{A,\mathbf{R}})}{\partial \mathbf{R}},
\end{equation}
$\partial \phi/\partial t$ can be replaced by $\mathbf{v}_{\mathbf{R}}$.
The rate of effective free energy becomes
\begin{eqnarray}
\dot{\tilde{A}} = \int \rm{d} \mathbf{R} \mathbf{v}_{\mathbf{R}} \cdot\left[\phi\nabla_{\mathbf{R}} \frac{\partial\tilde{\mathcal A}}{\partial \phi}\right].
\end{eqnarray}

The overall Rayleighian can be expressed as 
\begin{eqnarray}
\mathcal R/L_z &=& \Phi_{\mathbf{v}_{\mathbf{R}}}/L_z+\dot{\tilde{A}}/L_z \nonumber\\
&=& \int \rm{d} \mathbf{R}\left[ \frac{1}{2}\xi_\mathbf{R}\mathbf{v}^2_{A,\mathbf{R}} + \mathbf{v}_{\mathbf{R}} \cdot\left(\phi\nabla_{\mathbf{R}} \frac{\partial\tilde{\mathcal A}}{\partial \phi}\right)\right]. 
\end{eqnarray}
By minimizing this Rayleighian functional, one can obtain the equation of balance of force 
\begin{eqnarray}
\mathbf{v}_{\mathbf{R}} = -\frac{1}{\xi} \left[\phi\nabla_{\mathbf{R}} \frac{\partial\tilde{\mathcal A}}{\partial \phi}\right].
\end{eqnarray}
Combining with the conservation equation, Eq.~(\ref{conservation})
The Smoluchowski type of dynamic equation can be derived following the procedure of the Onsager principle.
\newpage 

\section{4.Thermodynamic theory of phase separation and interfacial tension in equilibrium systems} 
\label{subsec:chi_gamma_derivation}
The previously discussed content has addressed the phase boundaries and interfacial structures observed in simulations. On the theoretical level, the thermodynamics of phase boundaries and interfaces in equilibrium systems can be well described. In equilibrium, phase separation arises from the competition between the translational entropy of each component and the effective interactions between components\cite{klamserThermodynamicPhasesTwodimensional2018}. Translational entropy tends to promote mixing of different components, while the effective interactions between components drive the aggregation of like particles, overcoming the effect of translational entropy and thereby inducing phase separation\cite{siebertCriticalBehaviorActive2018}. For a binary system, phase separation in equilibrium can be described by the following free energy expression\cite{SoftMatterPhysics2013}:
$
f_0(\phi) = \phi \ln(\phi) + (1 - \phi) \ln(1 - \phi) + \chi \phi (1 - \phi),
$
Here, $\phi$ denotes the volume fraction of one component, and $\chi$ represents the effective interaction parameter between the components. Phase separation occurs when $\chi>\chi_c$
The spinodal boundary serves as a criterion for determining the stability limit of phase separation and is defined by the condition that the second derivative of the free energy density equals zero:$\frac{d^2 f}{d \phi^2} = 0$. Let us first compute the first derivative:
\begin{equation}
\frac{df}{d\phi} = \ln \phi - \ln(1 - \phi) + \chi (1 - 2\phi)\text{,}
\end{equation}
then compute the second derivative:
\begin{equation}
\frac{d^2 f}{d \phi^2} = \frac{1}{\phi} + \frac{1}{1 - \phi} - 2\chi\text{,}
\end{equation}
set it to zero to obtain the spinodal condition:
\begin{equation}
\frac{1}{\phi} + \frac{1}{1 - \phi} = 2\chi\text{,}
\end{equation}
after simplification, the expression for the spinodal line is obtained:
\begin{equation}
\chi_s(\phi) = \frac{1}{2} \left( \frac{1}{\phi} + \frac{1}{1 - \phi} \right)\text{,}
\end{equation}
we can also write it in the form of an inverse function by solving for $\phi$ in terms of $\chi$, yielding:
\begin{equation}
\phi = \frac{1}{2} \pm \frac{1}{2} \sqrt{1 - \frac{2}{\chi}}\text{.}
\end{equation}

The binodal represents the condition under which the system separates into two stable phases: equal chemical potentials and equal osmotic pressures, that is, satisfying:
\begin{equation}
\mu(\phi_1) = \mu(\phi_2), \quad f(\phi_1) - \mu \phi_1 = f(\phi_2) - \mu \phi_2\text{,}
\end{equation}
usually, this is obtained by constructing a common tangent. Let $\mu = df/d\phi$, then:
\begin{equation}
\mu = \ln \phi - \ln (1 - \phi) + \chi (1 - 2\phi)\text{.}
\end{equation}
then, solve for two points $\phi_1$ and $\phi_2$ that satisfy: $\mu(\phi_1) = \mu(\phi_2)$. Together, these constitute the method of constructing the common tangent, which is typically solved numerically. Specifically, for symmetric mixtures (e.g., around $\phi_0 = 0.5$), a simplified expression for the binodal can be derived. We obtain this using the common tangent condition:
\begin{equation}
\chi_b = \frac{1}{1 - 2\phi_0} \ln\left(\frac{1 - \phi_0}{\phi_0} \right)\text{.}
\end{equation}

A multiphase inhomogeneous system can be described by a Ginzburg–Landau form of the free energy,
\begin{equation}
F(\phi) = \int {\rm d} \mathbf{r}  [f_0 + c (\nabla \phi)^2],
\end{equation}
where $c$ is a constant. According to mean-field theory, the interfacial profile at equilibrium can be expressed as follows
\begin{equation}\label{interface}
\phi(y) = \pm \phi_0 \tanh \left( \frac{y - y_0}{\sqrt{2} \lambda} \right)\text{,}
\end{equation}
where $\lambda$ is the interfacial width. The effective interaction parameter and interfacial tension can be expressed as functions of the correlation length, $\lambda=(c/(2-\chi))^{1/2}$ and $\gamma = 2 \sqrt{2} c \phi_0^2/(3 \lambda)$\cite{chaikin1995principles}. The specific procedure is as follows:
The interface satisfies the steady-state condition, and the variation of the free energy functional obeys the Euler–Lagrange equation:
\begin{equation}
\frac{\delta F}{\delta \phi} = 0\text{,}
\end{equation}
by taking the variation with respect to $\phi(z)$, we obtain:
\begin{equation}
2c \frac{d^2 \phi}{dz^2} = \frac{d f_0}{d\phi}\text{,}
\end{equation}
calculate the derivative of $f_0(\phi)$:  
\begin{equation}
\frac{d f_0}{d\phi} = \ln \phi - \ln(1 - \phi) + \chi (1 - 2\phi)\text{,}
\end{equation}
therefore, the interfacial equation is:
\begin{equation}
2c \frac{d^2 \phi}{dz^2} = \ln \phi - \ln(1 - \phi) + \chi (1 - 2\phi)\text{.}
\end{equation}

Near the center of the interface $z_0$, the variation of $\phi$ is smooth. Let $\phi \approx 1/2$, denoted as $\phi = 1/2 + \delta\phi$, where $\delta\phi$ is a small deviation. Perform a second-order expansion of $f_0(\phi)$:
\begin{equation}
\ln \phi - \ln(1 - \phi) \approx 4\delta\phi, \quad \chi (1 - 2\phi) \approx -2\chi\delta\phi\text{,}
\end{equation}
substituting in, we obtain:
\begin{equation}
2c \frac{d^2 \delta\phi}{dz^2} = (4 - 2\chi)\delta\phi\text{.}
\end{equation}
This is a standard hyperbolic tangent equation, whose solution is:
\begin{equation}
\phi(z) = \frac{1}{1 + e^{-(z - z_0)/\lambda}}\text{,}
\end{equation}
the interfacial width $\lambda$ is given by:
\begin{equation}
\lambda = \sqrt{\frac{c}{2 - \chi}}\text{.}
\end{equation}

The definition of interfacial tension is:
\begin{equation}
\gamma = \int dz \left[ c \left(\frac{d\phi}{dz}\right)^2 \right]\text{,}
\end{equation}
using the derivative of $\phi(z)$:
\begin{equation}
\frac{d\phi}{dz} = \frac{\phi_0}{\sqrt{2\lambda}} \operatorname{sech}^2 \left( \frac{z - z_0}{\sqrt{2\lambda}} \right)\text{,}
\end{equation}
substituting it in:
\begin{equation}
\gamma = c \frac{\phi_0^2}{2\lambda} \int dz \, \operatorname{sech}^4 \left( \frac{z - z_0}{\sqrt{2\lambda}} \right)\text{,}
\end{equation}
using the integral formula:
$
\int_{-\infty}^\infty \operatorname{sech}^4(u) \, du = \frac{4}{3}\text{,}
$
finally, we obtain:
\begin{equation}
\gamma = \frac{2\sqrt{2} c \phi_0^2}{3\lambda}\text{.}
\end{equation}
The interface between two coexisting phases can be obtained from molecular simulations. By fitting the simulated interfacial profile using Eq.~(\ref{interface}), the interfacial width can be determined. Subsequently, based on the above relationships, the effective interaction parameter $\chi$ and surface tension $\gamma$ in the phenomenological free energy can be related to the microscopic interaction parameters.
\newpage
 
\section{5.Mean-field theoretical predictions of interfacial width
}
\label{Interfacefitting}
Based on the similarity between interfacial behaviors in nonequilibrium active matter systems and those in thermodynamic equilibrium systems, this study attempts to draw on the theoretical framework in mean-field theory regarding the relationship between interfacial energy and interfacial width to predict the trends of interfacial width and interfacial energy in active particle systems under nonequilibrium conditions.  
We demonstrate that the interfacial width can be used to predict the effective interaction parameter $\chi$ in the system. Furthermore, based on the predictions of the spinodal and binodal lines in mean-field theory, we construct a theoretical phase diagram and compare it with the phase separation critical points obtained directly from numerical simulations. If the theoretical binodal line shows good agreement with the numerically determined phase boundary, it indicates the validity of this analogy. This not only verifies the feasibility of extending equilibrium system theories to nonequilibrium active systems, but also provides a theoretical foundation and practical approach for predicting effective interfacial energy in nonequilibrium systems based on interfacial morphology.  

Under the condition $\Delta qE> (\Delta qE)_c$, we plot and analyze the steady-state interfacial morphologies corresponding to different values of $\Delta qE$. To improve statistical accuracy, we averaged the interfacial density profiles obtained under different external field strengths $E$ for the same $\Delta qE$. Figure~\ref{fig:fig3} shows the density profiles $\phi(y)$ along the direction perpendicular to the interface ($y$-direction) under different $\Delta qE$ conditions, where the blue squares represent simulation data and the orange solid lines represent fitted curves predicted by mean-field theory. It can be observed that as $\ DeltaqE$ increases, the interface evolves from being relatively smooth to increasingly steep. To quantitatively characterize the interfacial morphology, we fit the density profile $\phi(y)$ using a nonlinear least-squares method, employing the following hyperbolic tangent function as the fitting model:  
\begin{equation}
  \phi(y) = -\phi_0 \tanh \left( \frac{y - y_0}{\sqrt{2} \lambda} \right) + c
\end{equation}  
\begin{figure}[htbp]
  \centering
  \includegraphics[width=0.8\textwidth]{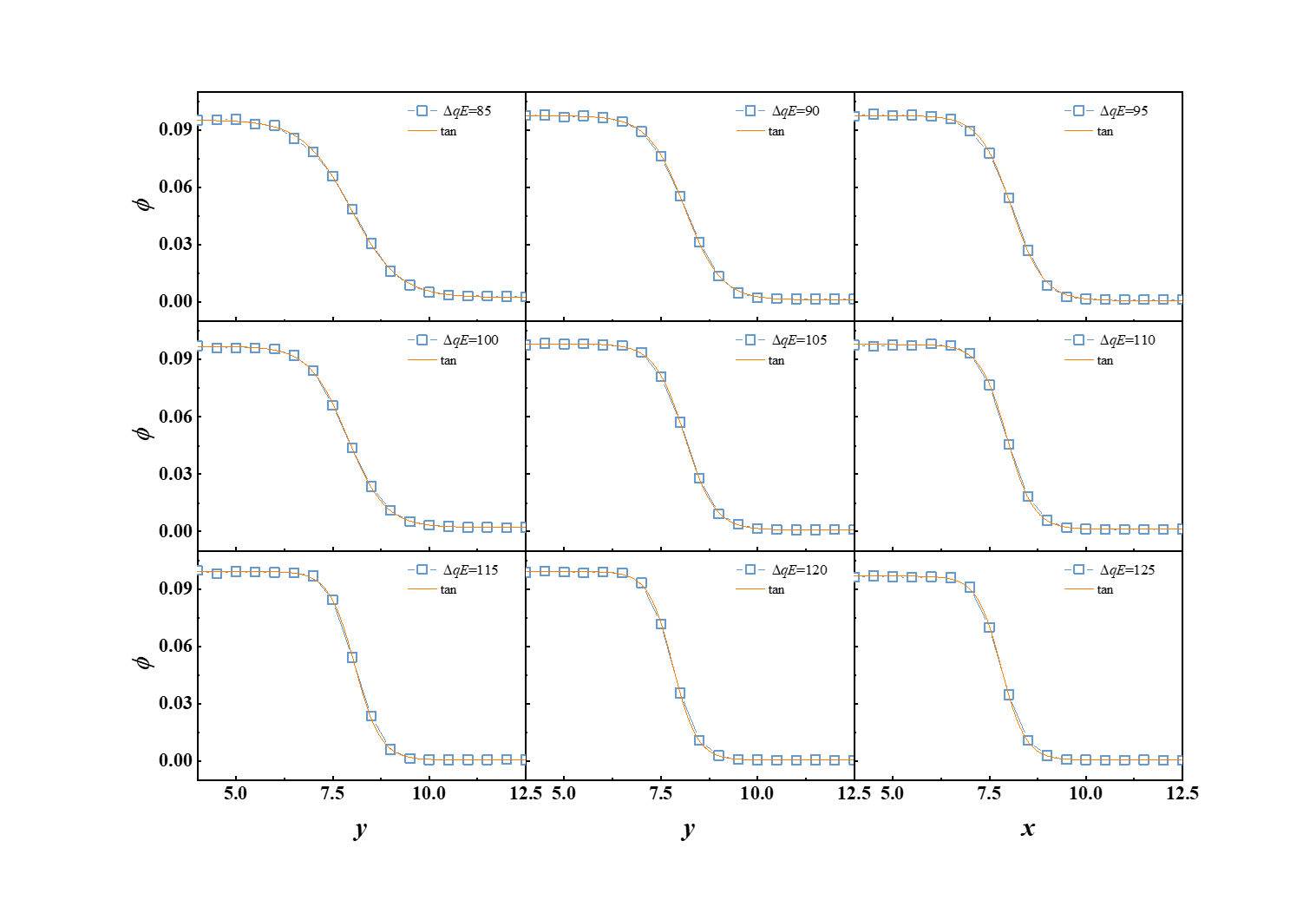}
  \caption{The density profiles $\phi(y)$ along the direction perpendicular to the interface under different $\Delta qE$, where the blue square symbols represent the numerical simulation results and the orange solid lines represent the mean-field fitting curves.}
  \label{fig:fig3}
\end{figure}
Through fitting the interfacial density profiles, we extracted the system's correlation length $\lambda$ and further analyzed its scaling relationship with $\Delta qE$. The results show that $\lambda$ satisfies the scaling relation $\lambda \sim (\Delta qE)^k$, where $k = -0.23$. This result indicates that with increasing external field strength, the interfacial width decreases, and the system tends to form sharper interfaces, making phase-separated structures more pronounced. Furthermore, based on mean-field theory, we derived the scaling relationship between the system’s effective interaction parameter $\chi$ and the surface tension $\gamma$. The parameter $\chi$ increases with $\Delta q E$ in a power-law form, satisfying $\chi \propto 2 - (\Delta q E)^{0.46}$, while the normalized surface tension $\gamma / c\phi_0^2$ exhibits a similar scaling behavior, $\gamma / c\phi_0^2 \propto (\Delta q E)^{0.23}$. Finally, we compare the theoretical results derived from mean-field theory—specifically, the binodal line represented by the blue curve and the spinodal line represented by the yellow curve in Fig. 1(d) of the main text—with the critical points of phase separation directly obtained from numerical simulations. The results show good agreement between the two. This indicates that, under nonequilibrium driving, spatial structural information such as interfacial width and density profiles can still be effectively described by drawing an analogy to mean-field theory in equilibrium systems, thereby allowing the prediction of effective interfacial energy and phase behavior. This provides reliable theoretical support and practical tools for understanding phase transition processes in active matter systems.
\newpage

\section{6.Partial-wetting phase} 
\subsection{Dynamic characteristics of the partial-wetting phase} 
\begin{figure}[htbp]
  \centering
  \includegraphics[width=0.5\textwidth]{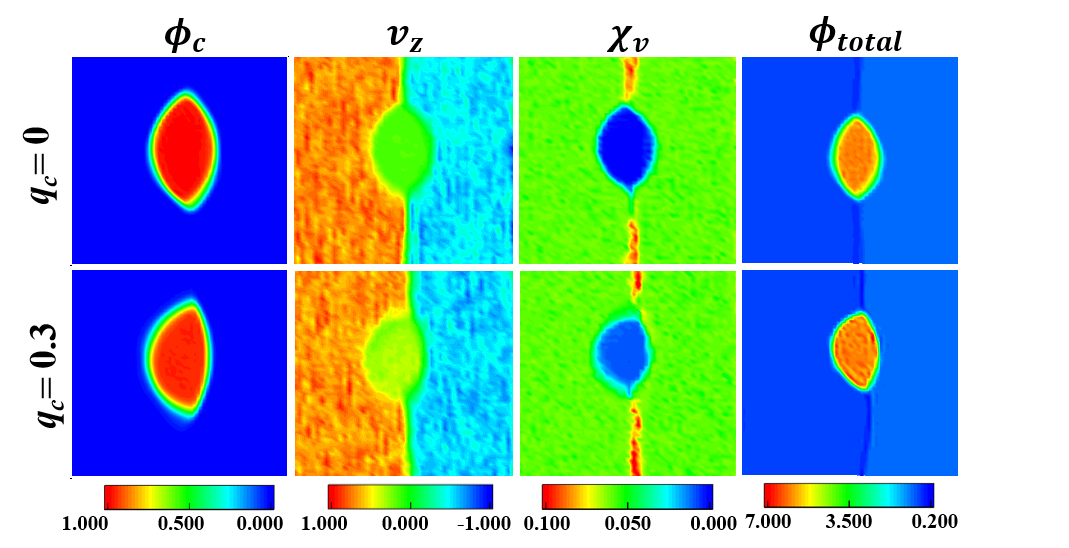}
  \caption{
    Heatmaps of the density distribution $ \phi_C $, velocity field in the $ z $-direction $ v_z $, velocity fluctuation $ \chi_v $, and the total density distribution $ \phi_{\text{total}} $ of all particles.
  }
  \label{fig:fig4}
\end{figure}

Figure~\ref{fig:fig4} presents the density distribution of C particles $ \phi_C $, the velocity field along the external field direction $ v_z $, the velocity fluctuation $ \chi_v $, and the total particle density distribution $ \phi_{\text{total}} $ projected onto the cross-section. Within the condensed phase, particles of the same species exhibit equal average drift velocities. Since components A and B experience equal but opposite driving forces, the drift velocities in the A-rich and B-rich phases are of equal magnitude but opposite direction. For $ q_C = 0 $, the drift velocity of the C condensed phase approaches zero. As $ q_C$ increases (e.g., $ q_C = 0.3$), the drift velocity becomes finite. A velocity gradient emerges at the three-phase interfaces, indicating that interfacial tension arises from these velocity variations. The velocity fluctuation $ \chi_v$ indicates that velocity fluctuations in the bulk phases of A and B are nonzero. Spatial fluctuations are also observed in the velocity $ v_z$ profile, which result from random interparticle collisions. The characteristic timescale of these fluctuations corresponds to the relaxation time along the external field direction. It is assumed to be a fast process within the phenomenological framework of this study. However, within the C condensed phase, the velocity fluctuations approach zero, indicating that particle motion is restricted by interfacial tension. The influence of interfacial tension can also be observed in the average particle density. This effect is further quantified by examining the total particle density distribution projected onto the cross-section,  
$
\phi_{\text{total}}(\mathbf{R}) = \sum_{i=1}^{n} \delta(\mathbf{R}_i - \mathbf{R}) / \rho_0,
$
where $ \rho_0 $ is the mean surface density of particles. The results indicate that within the C condensed phase, particle density is significantly increased due to interfacial tension, leading to higher internal pressure. This pressure effect suppresses drift velocity fluctuations. Additionally, at the A-B interface, the interfacial energy is maximized, leading to the highest collision probability between A and B particles. Consequently, velocity fluctuations at the A-B interface reach their maximum value.

\subsection{Van Hove correlation function analysis of particles in the partial-wetting phase} 
To further characterize the dynamical behavior of active particles in the direction perpendicular to the external field, we analyzed the van Hove correlation function of particles in the bulk phase and the partial-wetting condensed phase. The van Hove correlation function is defined as:  
$
G(\Delta r, \Delta t) = \langle \delta(\mathbf{r}(t + \Delta t) - \mathbf{r}(t) - \Delta r) \rangle,
$  
where $\Delta r$ represents the displacement of a particle over a time interval $\Delta t$, $\mathbf{r}(t)$ denotes the position of the particle at time $t$, and the averaging is performed over all particles. We computed the displacement distribution functions $G(\Delta y, \Delta t)$ and $G(\Delta x, \Delta t)$ along the $y$-axis (perpendicular to the external field, shown in the left column of the figure) and along the $x$-axis (parallel to the external field, shown in the right column), respectively.

\begin{figure}[htbp] 
  \centering
  \includegraphics[width=0.5\textwidth]{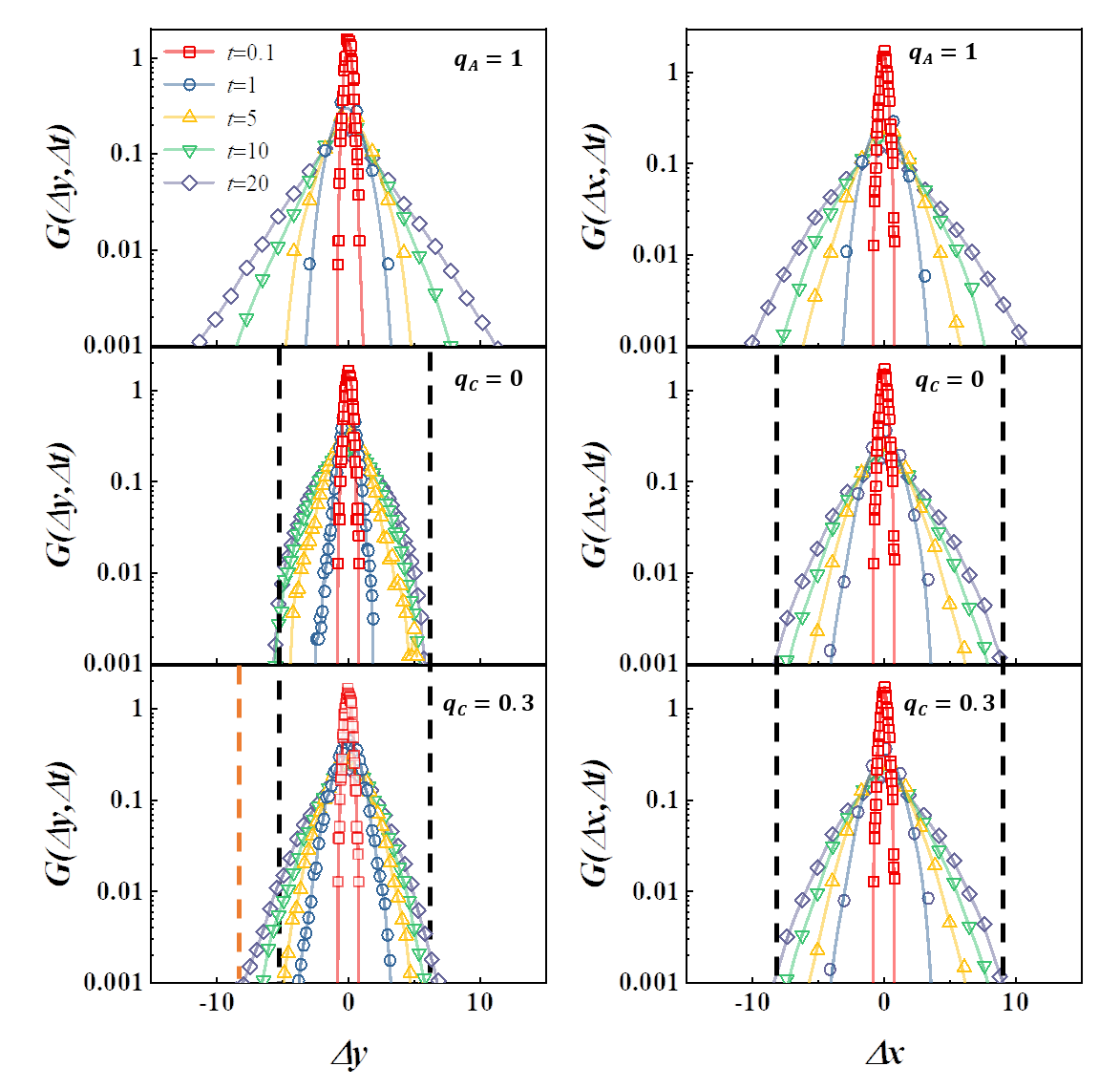}
  \caption{The van Hove correlation function $G(\Delta y, \Delta t)$ and $G(\Delta x, \Delta t)$ of particles along the $y$-direction and $x$-direction for $q_A = 1$ and different values of $q_C$. Different symbols correspond to different time intervals $\Delta t$.
  }
  \label{fig:fig5}
\end{figure}

Figure \ref{fig:fig5} shows the variations of $G(\Delta y, \Delta t)$ (left) and $G(\Delta x, \Delta t)$ (right) for C particles under fixed coupling coefficients $q_A = 1$ and $q_B = -1$ for A and B particles, respectively. Two cases, $q_C = 0$ and $q_C = 0.3$, are compared over a time scale ranging from $\Delta t = 0.1$ to $20$. When $q_C = 0$, C particles form stable condensed droplets at the AB interface and generate a pronounced interfacial tension, thereby suppressing their motion in the direction close to the interface (the $y$-direction). The van Hove function $G(\Delta y, \Delta t)$ decays rapidly at large displacements, indicating that the diffusion of C particles is significantly suppressed and that their motion in the vertical direction is strongly confined by the AB phase interface, being limited to the interior of the droplet. The black dashed line in the figure indicates the boundary of this confinement, further confirming the spatially restricted behavior.

When $q_C = 0.3$, asymmetric drift enhancement appears for C particles, and the confining effect of the interface formed by the A phase weakens. At this point, the van Hove distribution of C particles in the direction toward the A phase becomes significantly broader, especially in the extended tail of $G(\Delta y, \Delta t)$, indicating an increased range of motion in the vertical direction. In contrast, toward the B phase, C particles remain strongly compressed, with the displacement distribution remaining narrow, reflecting a pronounced asymmetric diffusion behavior. The external field coupling coefficient $q_C$ of C particles controls their “confinement” within the condensed phase and their interfacial diffusion capability. A smaller $q_C$ leads to strong interfacial confinement and more symmetric diffusion, whereas a larger $q_C$ reduces confinement on the phase side and results in a clearly directional diffusion behavior.
\newpage

\section{7.Contact angle variation in the asymmetric system under external field} 
\begin{figure}[htbp]
  \centering
  \includegraphics[width=0.8\textwidth]{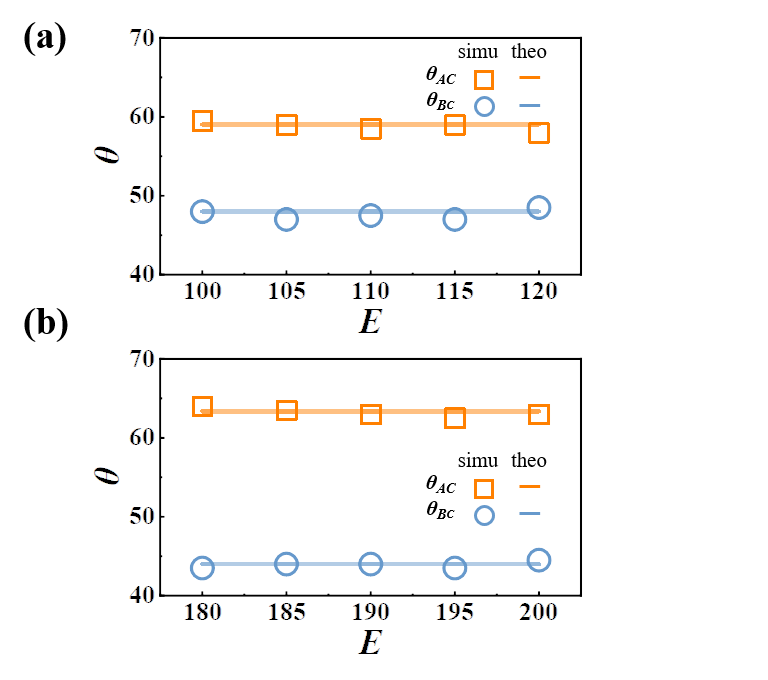}
  \caption{Variation of the contact angles $\theta_{AC}$ and $\theta_{BC}$ under different external field strengths $E$, (a) for $q_C = 0.3$ and (b) for $q_C = 0.5$. In the figure, the orange square symbols represent the simulation data of $\theta_{AC}$, and the blue circular symbols represent the simulation data of $\theta_{BC}$; the corresponding solid lines are the theoretical predictions calculated based on mean-field theory and the modified Young’s equation. The results show that the contact angles remain nearly constant with respect to $E$ and are mainly determined by $q_C$. Moreover, the asymmetry between $\theta_{AC}$ and $\theta_{BC}$ increases with increasing $q_C$.
  }
  \label{fig:fig6}
\end{figure}

In the main text, we have analyzed the variation of the contact angle under different external field strengths $E$ for the case of $q_C = 0$. Figure \ref{fig:fig6} further presents the contact angle behavior of the C condensed phase in an asymmetric three-phase system for $q_C = 0.3$ and $q_C = 0.5$. In the figure, the orange square symbols represent the simulation results of the contact angle $\theta_{AC}$ at the A–C interface, and the blue circular symbols represent the simulation results of the angle $\theta_{BC}$ at the B–C interface. The corresponding solid lines represent the theoretical values calculated based on a modified Young's equation, reflecting the predictions of mean-field theory.

At steady state, the three-phase contact point must satisfy the following two equilibrium conditions \cite{safranStatisticalThermodynamicsSurfaces2018}:
\begin{eqnarray}
\gamma_{AB} = \gamma_{AC} \cos(\theta_{AC}) + \gamma_{BC} \cos(\theta_{BC}) \nonumber
\\
\gamma_{AC} \sin(\theta_{AC}) = \gamma_{BC} \sin(\theta_{BC}).
\end{eqnarray}
Meanwhile, according to mean-field theory, the interfacial tension satisfies the following scaling relation with the microscopic driving force:
\begin{eqnarray}
\frac{\gamma_{\alpha\beta}}{c \phi_0^2} \propto (\Delta q_{\alpha\beta} E)^k,
\end{eqnarray}
where $\alpha, \beta = A, B, C$, and the exponent $k = 0.23$. The differences in coupling coefficients are defined as:
\begin{eqnarray}
\Delta q_{AB} = 2, \quad \Delta q_{AC} = 1 - q_C, \quad \Delta q_{BC} = 1 + q_C.
\end{eqnarray}

For $q_C = 0.3$, substituting the above relations yields the following equations for the contact angles:
\begin{eqnarray}\label{Young_q}
0.7^{0.23} \cos(\theta_{AC}) + 1.3^{0.23} \cos(\theta_{BC}) &=& 2^{0.23}, \nonumber \\
0.7^{0.23} \sin(\theta_{AC}) &=& 1.3^{0.23} \sin(\theta_{BC}).
\end{eqnarray}
The theoretical solution gives $\theta_{AC} \approx 59^\circ$, $\theta_{BC} \approx 48^\circ$.

For the case of stronger asymmetry with $q_C = 0.5$, the exponent terms in the above expressions are updated to: $\Delta q_{AC} = 0.5, \quad \Delta q_{BC} = 1.5$,
from which the theoretical contact angles are calculated as $\theta_{AC} \approx 63.34^\circ$, $\theta_{BC} \approx 43.96^\circ$.

As shown in Fig.~\ref{fig:fig6}, in both $q_C$ cases, the contact angles $\theta$ in the simulation results exhibit strong stability with respect to variations in the external field strength $E$, indicating that the contact angle is mainly controlled by the coupling coefficient $q_C$ of component C and is insensitive to changes in $E$. Moreover, as $q_C$ increases, the wettability of C particles toward phases A and B becomes increasingly asymmetric: the contact area at the A–C interface gradually increases, while that at the B–C interface gradually decreases, indicating that the interfacial morphology is regulated by the charge–field coupling strength. Furthermore, with increasing $q_C$, the asymmetry of the contact angles continues to grow, reflecting a selective wetting effect of the C condensed phase under nonequilibrium driving. These results demonstrate that the charge coupling parameter $q_C$ can serve as a key control parameter for regulating wetting behavior in nonequilibrium systems. The theoretical predictions agree well with the simulation results for both values of $q_C$, further confirming the validity of mean-field theory in describing interfacial tension and wetting behavior in nonequilibrium active matter systems.

\end{document}